\documentclass[12pt]{article}

\usepackage{newtxtext,newtxmath}

\usepackage{graphicx}

\usepackage[letterpaper,margin=1in]{geometry}

\renewenvironment{abstract}
	{\quotation}
	{\endquotation}

\date{}

\makeatletter
\renewcommand{\fnum@figure}{\textbf{Figure \thefigure}}
\renewcommand{\fnum@table}{\textbf{Table \thetable}}
\makeatother

\usepackage{scicite}

\usepackage{url}

\def\scititle{
	System of Agentic AI for the Discovery of Metal-Organic Frameworks
}
\title{\bfseries \boldmath \scititle}

\author{%
Théo~Jaffrelot~Inizan$^{1,2,3\ast}$, Sherry~Yang$^{4,5\dagger}$, Aaron~Kaplan$^{3\dagger}$, \\
Yen-hsu~Lin$^{1,6\dagger}$, Jian~Yin$^{1,6}$, Saber~Mirzaei$^{1,6}$, \\
Mona~Abdelgaid$^{2,3}$, Ali~H.~Alawadhi$^{1,6}$, KwangHwan~Cho$^{3}$, \\
Zhiling~Zheng$^{1,6}$, Ekin~Dogus~Cubuk$^{5}$, Christian~Borgs$^{2,4}$, \\
Jennifer~T.~Chayes$^{2,4,7,8,9}$, Kristin~A.~Persson$^{2,3,10\ast}$, Omar~M.~Yaghi$^{1,2,6\ast}$\and
\small$^{1}$Department of Chemistry, University of California, Berkeley, CA, USA.\and
\small$^{2}$Bakar Institute of Digital Materials for the Planet, University of California, Berkeley, CA, USA.\and
\small$^{3}$Materials Sciences Division, Lawrence Berkeley National Laboratory, Berkeley, CA, USA.\and
\small$^{4}$Department of Electrical Engineering and Computer Sciences, University of California, Berkeley, CA, USA.\and
\small$^{5}$Google DeepMind, Mountain View, CA, USA.\and
\small$^{6}$Kavli Energy NanoScience Institute, Berkeley, CA, USA.\and
\small$^{7}$Department of Mathematics, University of California, Berkeley, CA, USA.\and
\small$^{8}$Department of Statistics, University of California, Berkeley, CA, USA.\and
\small$^{9}$School of Information, University of California, Berkeley, CA, USA.\and
\small$^{10}$Department of Materials Science and Engineering, University of California, Berkeley, CA, USA.}

\begin{document} 

\maketitle

\begin{abstract} \bfseries \boldmath
Generative models and machine learning promise accelerated material discovery in MOFs for CO$_{2}$ capture and water harvesting but face significant challenges navigating vast chemical spaces while ensuring synthetizability. Here, we present MOFGen, a system of Agentic AI comprising interconnected agents: a large language model that proposes novel MOF compositions, a diffusion model that generates crystal structures, quantum mechanical agents that optimize and filter candidates, and synthetic-feasibility agents guided by expert rules and machine learning. Trained on all experimentally reported MOFs and computational databases, MOFGen generated hundreds of thousands of novel MOF structures and synthesizable organic linkers. Our methodology was validated through high-throughput experiments and the successful synthesis of five``AI-dreamt'' MOFs, representing a major step toward automated synthesizable material discovery.

\end{abstract}

\noindent
Accelerating the discovery of novel, synthesizable materials is crucial for addressing major global challenges, including carbon capture \cite{lin_scalable_2021,zhou_carbon_2024,rohde_high-temperature_2024}, water harvesting \cite{rieth_record_2017,towsif_abtab_reticular_2018,hanikel_evolution_2021,zheng_high-yield_2023}. Historically, materials discovery has heavily relied on trial-and-error approaches, experimental heuristics, incremental improvements, chemical intuition, and computational simulations, notably with density functional theory (DFT) \cite{hohenberg_inhomogeneous_1964,kohn_self-consistent_1965}. Significant advances in the development of open-access materials databases \cite{jain_commentary_2013,ghahremanpour_alexandria_2018,sriram_open_2024,rosen_machine_2021,chung_advances_2019,zhao_core_2024}, automated laboratory techniques \cite{szymanski_autonomous_2023, rao_machine_2022,domingues_using_2022}, machine-learning-based property predictors, generative models \cite{zeni_generative_2025,yang_scalable_2024,fu_mofdiff_2023} and machine learning force fields (MLFF) \cite{behler_generalized_2007,smith_ani-1_2017,schutt_schnet_2018,wang_deepmd-kit_2018,batzner_e3-equivariant_2022,inizan_scalable_2023,ple_force-field-enhanced_2023,batatia_mace_2023,batatia_foundation_2024} have collectively enabled exploration of chemical spaces and generated large databases of millions of computationally stable candidates \cite{merchant_scaling_2023, sriram_open_2024}.
However, synthesizability remains largely unexplored and presents a major limitation in translating predictions to reality. Indeed, a material that appears stable from a quantum mechanical perspective can be extremely difficult or too costly to synthesize.

Metal–organic frameworks (MOFs) \cite{yaghi_reticular_2003,furukawa_chemistry_2013,hanikel_evolution_2021}, nanoporous materials formed from metal nodes linked by strong bonds to organic molecules, are an example of materials synthesized in crystalline form.
Their synthesis typically requires extensive screening of experimental conditions (e.g., metal and linker concentrations, temperature) and involves multi-step organic linker syntheses, rendering the process labor-intensive and limiting exploration of the chemical space. To overcome these challenges, integrating generative models with quantum mechanical predictions, and rigorous experimental constraints specific to MOFs can streamline candidate selection and reduce reliance on conventional heuristics \cite{zheng_large_2025,fu_mofdiff_2023,park_generative_2024}. 

Recent breakthroughs in generative models, exemplified by AlphaFold in protein structure prediction \cite{jumper_highly_2021, abramson_accurate_2024} and diffusion-based models in inorganic crystal generation \cite{yang_scalable_2024,yang_generative_2024,zeni_generative_2025}, highlight the potential of these models to discover structures beyond human intuition. However, even if these models can generate a large number of DFT-stable materials, translating these advances to MOFs remains a challenge. For example, conventional stability metrics \cite{bartel_review_2022} are not always applicable to MOFs. In addition, MOFs typically involve large complex structures, chemical composition diversity, and a wide range of topologies, not trivial to capture by machine learning models. As a result, only a handful of inorganic material discovered by generative models has been experimentally verified to date \cite{zeni_generative_2025}, emphasizing the significant gap between computational prediction and real synthesis. 

This report outlines the design of a system of Agentic AI, MOFGen, combining large language models (LLMs) \cite{openai_gpt-4_2024,touvron_llama_2023}, diffusion algorithms \cite{sohl-dickstein_deep_2015,ho_denoising_2020,yang_generative_2024}, MLFFs, quantum mechanical computations \cite{perdew_generalized_1996,furness_accurate_2020}, machine learning-based synthesis predictors \cite{coley_scscore_2018,nandy_using_2021} and high-throughput synthesis. Trained across all reported MOF chemistry, our interconnected system of specialized agents enabled rapid exploration and discovery of synthesizable MOFs, significantly shortening the path from computational predictions to realization in the laboratory. Within just a few weeks, we have synthesized five ``AI-dreamt'' MOFs, with some completed in just few days. By leveraging generative models and AI-driven methodologies, our system accelerates the creation of ``AI-dreamt'' synthesizable materials, paving the way towards human-in-the-loop foundation model for MOF discovery. 

\section*{Results}

\subsection*{Overview of the system of Agentic AI, MOFGen}\label{agent_overview}
\begin{figure}
    \centering
    \includegraphics[width=\textwidth]{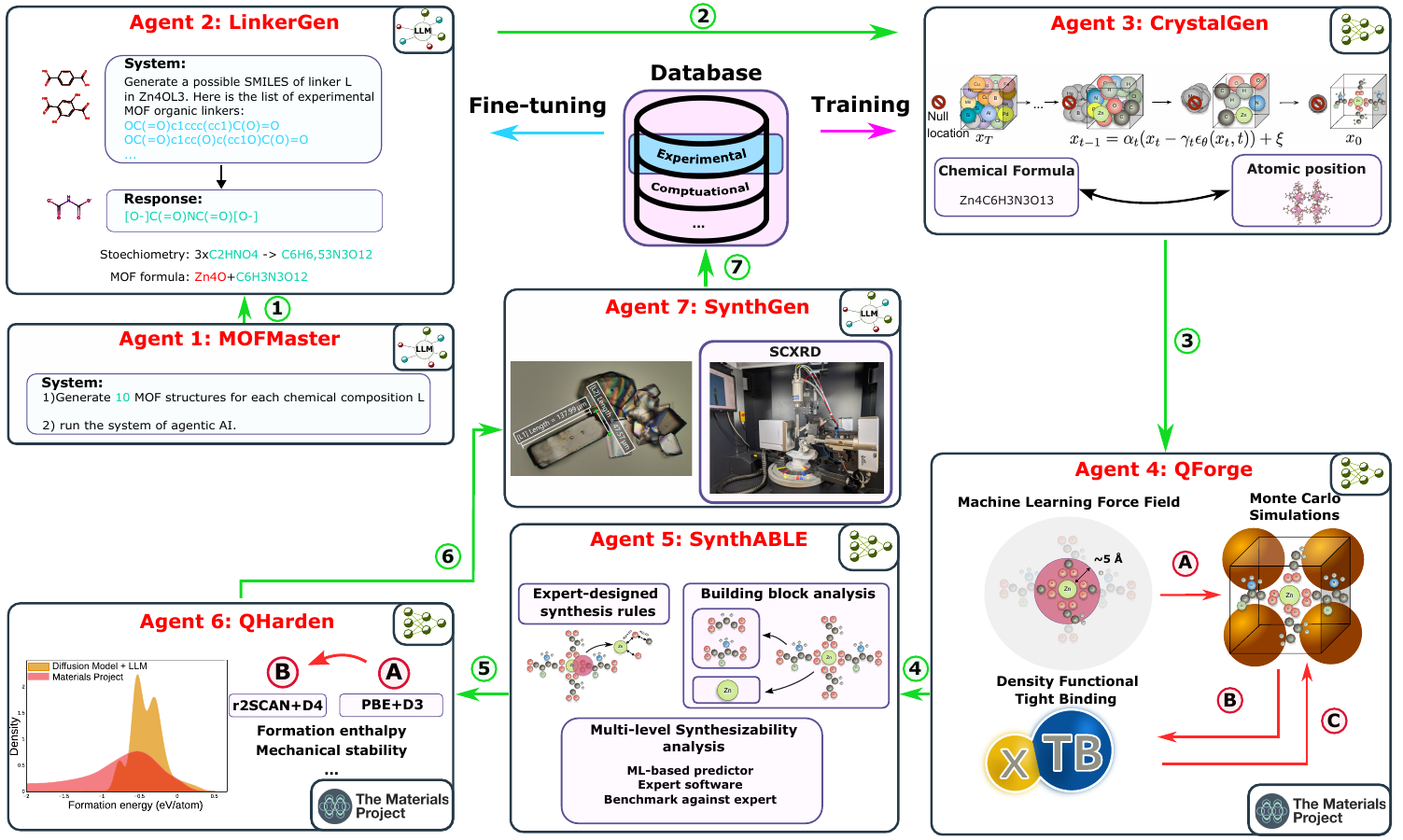}
    \caption{\textbf{Overview of MOFGen.} 
    1) \textit{MOFMaster}: Manages the overall system and serves as user-interface. 
    2) \textit{LinkerGen}: An in-context learning LLM agent that proposes novel chemical compositions. 
    3) \textit{CrystalGen}: A denoising diffusion probabilistic model conditioned on chemical compositions that generates crystal structures. 
    4) \textit{QForge}: sequence of geometry optimization and Monte Carlo sampling to filter non-porous structures
    5) \textit{SynthABLE}: Decomposes the diffusion-generated MOFs into their building blocks and assesses synthesizability using multi-fidelity rules and ML-based predictors. 
    6) \textit{QHarden}: sequence of medium-to-high level quantum mechanical geometry optimization and formation energy evaluation, from PBE-D4 to r$^2$SCAN-D4. 
    7) \textit{SynthGen}: High-throughput synthesis platform to synthesis MOF, combined with LLM synthesis planning system, the crystal are then passed through SCXRD or PXRD for charaterization and added to the pool of experimental structures.}
    \label{fig:figure1}
\end{figure}

Our approach, MOFGen, consists of a modular system of AI and computational agents, each defined as an entity capable of taking actions based on its environment to accelerate MOF discovery. The architecture of our system, shown in Fig.~\ref{fig:figure1}, is composed as follows. First, the main LLM agent, \emph{MOFMaster}, provides three generation options: (i) generate a specified number of MOF structures without chemical information, (ii) generate MOF structures containing a specific metal secondary building unit (SBU), and (iii) generate MOF structures with a given chemical composition. Second, these instructions are then passed to \emph{LinkerGen}, an LLM chemical-composition generator leveraging in-context learning and chain-of-thought \cite{wei_chain--thought_2023} on experimental MOF organic linkers, and which produces MOF chemical formulae based on the constraints provided by \emph{MOFMaster}.

Third, the generated formulae are served as inputs to \emph{CrystalGen} which is a chemical composition–conditioned, all-atom, denoising-diffusion probabilistic model for crystal structure generation, trained on both experimental and computational MOF crystal structures. Fourth, the diffusion-generated structures are subjected to a series of geometry optimization and filtering steps through a low-level quantum mechanical agent, \emph{QForge}. Fifth, \emph{SynthABLE} decomposes each MOF into its building blocks and evaluates their synthesizability using multi-level fidelity synthetic rules and machine learning synthesizability predictors. Sixth, the remaining candidates are further refined by a high-level quantum mechanical agent, \emph{QHarden}, which performs successive geometry optimizations ranging from PBE-D4 to the higher-level r$^2$SCAN-D4 levels. Seventh, \emph{SynthGen} ranks the most likely-to-be-synthesized MOF candidates, which are subsequently validated through our high-throughput synthesis platform, powered by LLM–guided synthesis planning and characterized using single crystal or powder X-ray diffraction (SCXRD or PXRD). 

\subsection*{Chemical Formula Generation with \emph{LinkerGen}}\label{LinkerGen_agent}
A dedicated LLM agent, \emph{LinkerGen}, is employed to generate organic linker chemical formulae. The agent is prompted with chemical representations extracted from the experimental databases through an in-context learning strategy using chemical constraints with desired metal SBU formula. We tested several strategies: (i) chemical formula $\rightarrow$ SMILES, and (ii) SMILES $\rightarrow$ SMILES. Our analysis shows that the SMILES-to-SMILES strategy exhibits the highest success rate of generating valid linkers. In addition, the synthesizability scores (SA score and SCScore) show (see the supplementary materials, section \ref{SI:Linker_Gen}) a shift in the distribution with a double peak matching the experimental data. Once generated the organic linker formulae are then concatenated with metal SBU formulae, using the correct stoichiometric ratios. To enhance structural complexity, supercell formulations are also generated by scaling the number of total atoms, linkers and metal SBUs, with a maximum of 256 atoms, atom limit that \emph{CrystalGen} can target. The resulting database of chemical formulae serves as the input for downstream diffusion-based structure generation for \emph{CrystalGen}. 

\subsection*{Crystal Structure Generation with \emph{CrystalGen}}\label{CrystalGen_agent}
Due to their large size and complex three-dimensional frameworks, MOFs present a significantly greater challenge for diffusion models compared to typical inorganic crystals, which usually contain fewer than a dozen atoms per unit cell. To effectively manage this complexity, we represent each MOF structure as a point cloud \cite{yang_generative_2024} (see the supplementary materials, section \ref{SI:diffusion}). Specifically, a MOF containing $A$ atoms is encoded into a matrix of shape $[A,4]$, where the first three columns represent the fractional coordinates and the fourth column encodes the atomic number. \emph{CrystalGen} initializes the structure using 256 atoms with random atomic number and random locations. During the denoising diffusion process, \emph{CrystalGen} removes irrelevant atoms and moves relevant atoms to probable locations, under the formulae guidance through classifier-free guidance.
\emph{CrystalGen} was trained on most of experimental and computational MOF databases (see the supplementary materials, section \ref{databases}), enabling the generation of MOF crystal structures with almost all synthetically-accessible metal SBUs and organic linkers chemical compositions. In this study, we focused on zinc-based MOFs due to their widespread application, and proven suitability for large-scale production \cite{noauthor_largescale_nodate, paul_scale-up_2023}. However, their popularity makes the discovery of novel structures even more challenging. Given a chemical formula proposed by \emph{LinkerGen} using zinc SBUs (Zn$_{4}$O), \emph{CrystalGen} generates ten candidate structures per formula, resulting in a database of 259,559 predicted MOF structures. The size of this database is comparable to other state-of-the-art generative databases for inorganic materials, underscoring the effectiveness of point-cloud-based approaches in capturing the geometric and chemical diversity inherent to MOFs.

\subsection*{Pre-Screening and Filtering with \emph{QForge}}\label{QForge_agent}
Given that structures generated by diffusion models deviate from equilibrium, a low-level quantum mechanical agent, \emph{QForge}, is employed for pre-screening. First, candidates are processed with Zeo++ \cite{willems_algorithms_2012}, under a set of expert-designed assumptions that a valid MOF must exhibit (see the supplementary materials, section \ref{SI:qforge}). The structures then undergo geometry optimization using the MACE-MP-0 universal MLFF, whose main advantage is computational efficiency, enabling rapid optimization even far from equilibrium \cite{batatia_foundation_2024}. Since MACE-MP-0 has so far been validated primarily for static properties, its direct application to MOF structure optimization is non-trivial. Nevertheless, it successfully optimized 251,118 candidate structures. This step quickly brings most structures close to equilibrium, effectively eliminating chemically unreasonable structures. An additional screening step with Zeo++, using identical criteria, retains 72$\%$ of these structures. Finally, geometry optimization is performed using the physics-based GFN1-xTB method, resulting in more reliable structures. Neither MACE-MP-0 nor GFN1-xTB has previously been applied at large scale for MOF relaxation, notably MACE-MP-0 has been mainly used for static or small-scale dynamical computations \cite{elena_machine_2024, lim_accelerating_2024}. Here, we demonstrate that the sequential application of both methods yields reliable geometrical structures that could be used in-tandem with diffusion models or high-throughput screening for MOFs.

\subsection*{Synthesizability Assessment with \emph{SynthABLE}}\label{SynthABLE_agent}
\begin{figure}
    \centering
    \includegraphics[width=\textwidth]{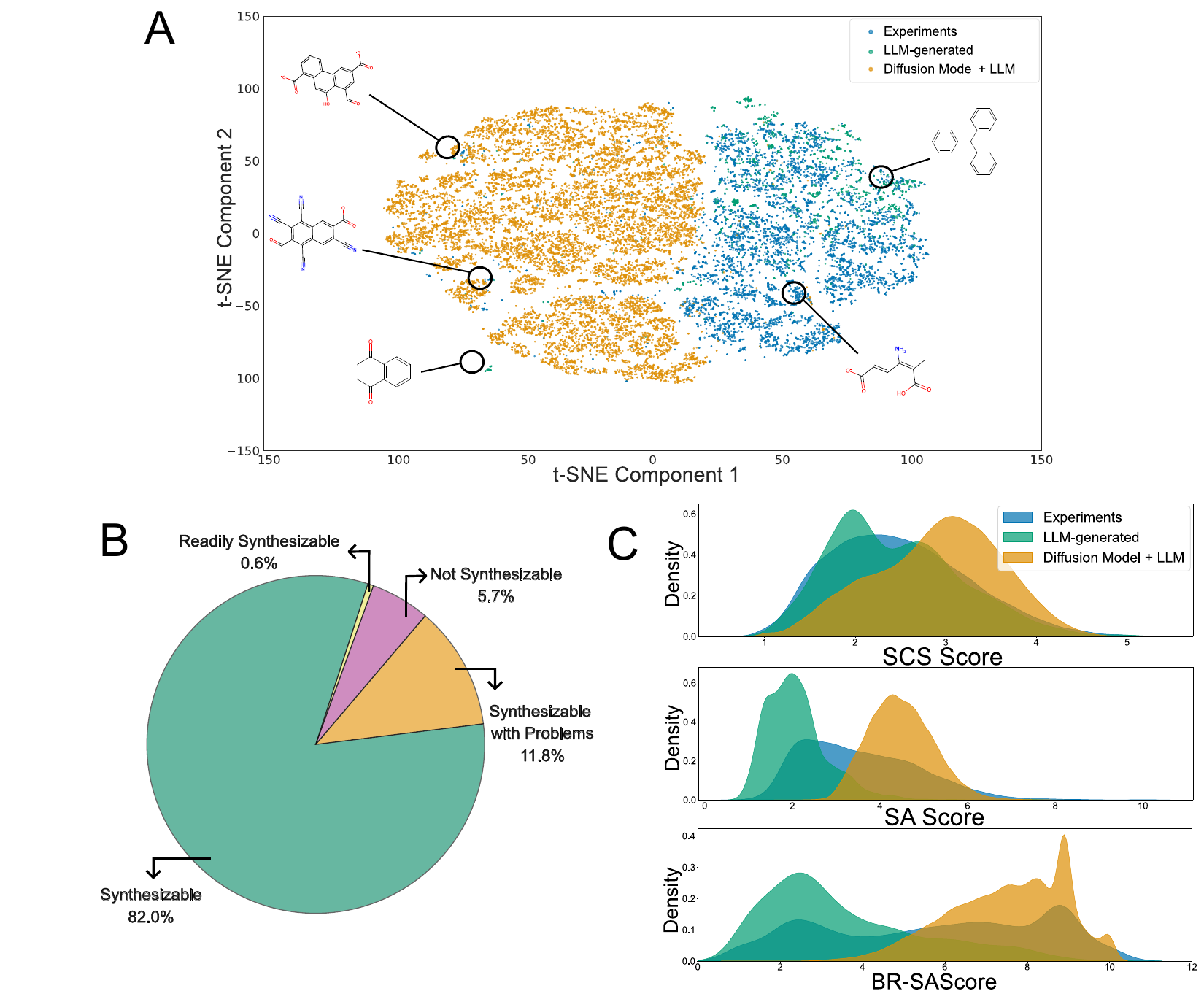}
    \caption{\textbf{Analysis of MOF's organic linkers with \emph{SynthABLE}.}
    Analysis of the 80,968 organic linkers extracted from the diffusion-generated MOF crystal structures using MOFid \cite{bucior_identification_2019}, after optimization and filtering through \emph{QForge}.
    \textbf{a}, t-SNE projection \cite{maaten_visualizing_2008} of diffusion-generated MOF linkers. For each molecule, the mean and standard deviation of atomic descriptors from the MACE-OMAT-0 model were computed and concatenated, which served as input to the t-SNE algorithm.  
    \textbf{b}, Synthesizability assessment with the Allchemy software. 
    \textbf{c}, SCScore \cite{coley_scscore_2018}, SA score \cite{ertl_estimation_2009} and BR-SAScore score distributions of the diffusion-generated organic linkers (Diffusion Model + LLM), compared to the curated experimental database and the initial LLM-generated linkers chemical formula used as input to the diffusion model.}
    \label{fig:figure2}
\end{figure}

Predicting the synthesizability of a MOF is inherently challenging and lacks a universal solution. One approach is to assess the synthesizability of individual components, such as organic linkers. After structures pass \emph{QHarden}, each MOF is decomposed into its building blocks, which are subsequently fed to \emph{SynthABLE}. This process results in approximately 180,000 novel and unique organic linkers. Initially, our agent leverages ML-driven predictions from the Allchemy software \cite{wolos_computer-designed_2022, zadlo-dobrowolska_computational_2024, strieth-kalthoff_artificial_2024}, indicating that 93.8$\%$ of a randomly selected batch of around 126,000 linkers are synthesizable or at least synthesizable with moderate difficulty (Fig.~\ref{fig:figure2}b). To evaluate this prediction, we implement a human-in-the-loop approach through a blind survey involving experimentalists. Each experimentalist independently evaluated a set of linkers predicted by the ML algorithm to be either synthesizable or non-synthesizable. We found that the experimentalists consistently identified the non-synthesizable linkers but were more cautious about the synthesizable ones, with success rate of 75 $\%$ and 40$\%$ respectively. These results may imply that the ML predictions could be overly optimistic regarding the synthesizability of linkers. Nonetheless, even under a conservative assumption that only 40$\%$ of the generated linkers are truly synthesizable, there would still be approximately 50,000 viable linkers, a number that greatly surpasses the number of  reported linkers to date \cite{groom_cambridge_2016}. Furthermore, we focused on zinc-based MOFs with reported SBUs (e.g., paddlewheel, tetrahedral motifs) for ease of synthesis, narrowing the candidate set to 80,000 linkers. Synthesizability is further quantified using scoring functions, including SA scores, SCScore (trained on the Reaxys database), SMILES length and the newly developed BR-SAScore \cite{chen_estimating_2024} (Fig.~\ref{fig:figure2}c) that show that the generated linkers exhibit synthesizability scores comparable with experiment. Compared to organic linkers generated by LLM-based methods (Fig.~\ref{fig:figure2}a), our diffusion-based approach yields chemical distribution that not overlap experimental data but also span a much broader chemical space, while preserving synthesizability through synthesis scores. Moreover, our model tends to generate longer organic linkers, resulting in structures with larger pore sizes, desirable for many MOF applications. Lastly, applying very strict expert design filters (see the supplementary materials, section \ref{SI:synthable}) we ranked the organic linkers based on their likely-to-be-synthesized basis, which reduces the set to 349 high-confidence organic linkers, which are subsequently selected by experimentalists for high-throughput synthesis.

\subsection*{Final Optimization with \emph{QHarden}}\label{QHarden_agent}
The top MOF candidates possessing synthesizable linkers are then refined by the high-level quantum mechanical agent, \emph{QHarden}. A multi-step cleaning process first verifies that interatomic distances match experimental values. Floating atoms, molecular fragments, and other artifacts are removed from the crystal structures. The structures then undergo a two-stage DFT optimization: an initial coarse geometry optimization using the PBE functional with D4 dispersion-correction, followed by a tight optimization with r$^2$SCAN-D4 (see supplementary materials, section \ref{SI:DFT_wf}). Although r$^2$SCAN has not been applied to MOF systems due to computational cost, our workflow produces a final database of 3,000 DFT-optimized MOF structures, accumulated over 30,000 DFT ionic steps. This constitutes one of the largest and most accurate MOF databases, thus enabling further use in the development of the next generation of MLFF for MOFs. We further assessed the mechanical and thermal stability of our DFT-optimized MOFs using machine-learning predictors for decomposition temperature and solvent-removal stability \cite{nandy_using_2021}. As shown in Fig.~\ref{fig:figure3}a, our structures exhibit a sharp decomposition temperature peak around 341\,K with a small standard deviation, closely matching (or slightly improved) experimental averages. The stability upon solvent removal is similarly enhanced (average value of 0.88 compared to 0.65 in the QMOF database; Fig.~\ref{fig:figure3}b). 

To further assess the mechanical stability we computed the bulk moduli with the MACE-MP-0 model (Fig.~\ref{fig:figure3}c). This represents the first large-scale application of MACE-MP-0 for bulk modulus evaluation, providing a computationally efficient alternative to DFT computations (see the supplementary materials, section \ref{SI:mace-mp0_analysis}). Our analysis reveals that the model has tendency to generates structures with small bulk moduli, indicative of enhanced dynamical flexibility or “breathing” behavior \cite{barthelet_breathing_2002, loiseau_rationale_2004, serre_explanation_2007}.
This dynamic pore adjustment can optimize guest–host interactions, improving selectivity in gas storage and separation (e.g., CO$_{2}$, CH$_{2}$, H$_{2}$), catalysis, sensing (pressure sensors and actuators), and controlled drug release. \cite{alhamami_review_2014} Furthermore, formation energy calculations show that diffusion-generated zinc-based MOFs have close distribution with the Material Project data (Fig.~\ref{fig:figure3}d), while having different chemical composition. To evaluate the predictive capability and generalizability of MOFGen, we also compared it with diffusion-generated MOF but with other metal SBUs (see the supplementary materials, section \ref{SI:formation_energies}). Most of the generated MOFs exhibit negative formation energies, suggesting physical-stability. In summary, all of this criteria thermal, chemical, physical and mechanical metrics underscore the efficacy of our agent-based architecture in generating stable, synthesizable MOF structures with various topologies and chemistry (Fig.~\ref{fig:figure3}d). 

\section*{Experimental validation with \emph{SynthGen}}\label{SynthGen_agent}\

From an initial set of 349 high-confidence linkers predicted by our system, we experimentally validated five of ``AI-dreamt'' MOFs, referred to as the AI-MOF series (see supplementary materials, section \ref{SI:material_method_synth}), following a three conceptual synthesis strategies: (i) crossover mutation synthesis, where previously synthesized organic linkers are combined with different metal SBUs (e.g., from aluminim to zinc) to explore novel geometries and applications; (ii) Reimagination synthesis, analogous to creative arts and image generation, in which organic linkers serve as inspiration and undergo subtle modifications by expert to enhance synthesizability or target performance; and (iii) \textit{de novo} synthesis, directly synthesizing MOF with ``AI-dreamt'' novel linker designs (Fig.~\ref{fig:figure4}). The synthesis conditions were systematically screened and optimized using an automated robotic platform, enabling efficient high-throughput experimentation by optimizing parameters such as metal-to-linker ratio, concentration of starting materials, and temperature.

\subsection*{Crossover mutation synthesis}
We find that MOFGen re-discovered organic linkers synthesized in MOFs with other metal SBUs than zinc. For example, the muconate was already reported with aluminium \cite{ntep_designing_2019}. This finding enables us to identify MOFs for applications that differ from their original purpose while accelerating their synthesis. The crystals of AI-MOF-1 were grown through solvothermal method\cite{mof5_2005,qiaowei_2009}. The experimental PXRD confirmed that AI-MOF-1 was successfully synthesized (Fig.~\ref{fig:figure4}a). While muconate can adopt either \textit{anti} or \textit{syn} conformations, the synthesized AI-MOF-1 exhibited the \textit{syn} conformation.

\subsection*{Reimagination synthesis}

Inspired by art and image generation, we introduce a strategy termed \emph{reimagination synthesis}, in which AI-generated organic linkers serve as inspiration, then refined by chemists through slight structural modifications, like a sketch. For example, one of the diffusion-generated linkers (Figure~\ref{fig:figure4}b), chrysene~—~a polycyclic aromatic hydrocarbon (PAH) that is challenging to synthesize~—~was modified from five to four-ring PAH to form commercially available perylene. Through reaction with zinc nitrate and using solvothermal synthesis, red, block-shaped crystals (AI-MOF-2) were obtained (see supplementary materials, section \ref{SI:synth_AI-MOF-2}). Experimental PXRD patterns matched the simulated structure. This represents the first reported MOF incorporating this particular PAH, not targeted previously because of low solubility due to $\pi$-$\pi$ stacking aggregation. 

\subsection*{\textit{de-novo} synthesis}
To  further demonstrate the predictive and discovery capabilities of MOFGen, we employed a \textit{de-novo} synthesis strategy, synthesizing MOFs from synthetically unreported organic linkers (see supplementary materials, section \ref{SI:synth_AI-MOF-3}-\ref{SI:synth_AI-MOF-7}). The linker 2,3-dimethyl-2-butenedioic acid was chosen for its high synthesizability and its frequent generation by our models. Using microwave-assisted synthesis with \textit{N, N}-Diethylformamide (DEF) as solvent, we successfully produced colorless crystalline materials (AI-MOF-3), whose PXRD patterns revealed orthorhombic space group.

We also performed an additional experiment using dimethylformamide (DMF) solvent and using the solvothermal strategy, successfully growing single crystals (AI-MOF-4). SCXRD revealed that two DMF molecules coordinate to the zinc SBUs distorting the symmetry from cubic to orthorhombic. This coordination mode is unique, since it has not been reported for zinc SBUs. This coordination flexbility around the Zn$^{2+}$ ions has been investigated in MOF-5 using different spectroscopic techniques \cite{brozek_dynamic_2015}. However, it has not been observed in the SCXRD structure. In addition, removing the coordinated DMF could create open metal sites beneficial for hydrogen storage and catalytic applications.\cite{dzubak_ab_2012, wei_metalorganic_2020}

\begin{figure}
    \centering
    \includegraphics[width=\textwidth]{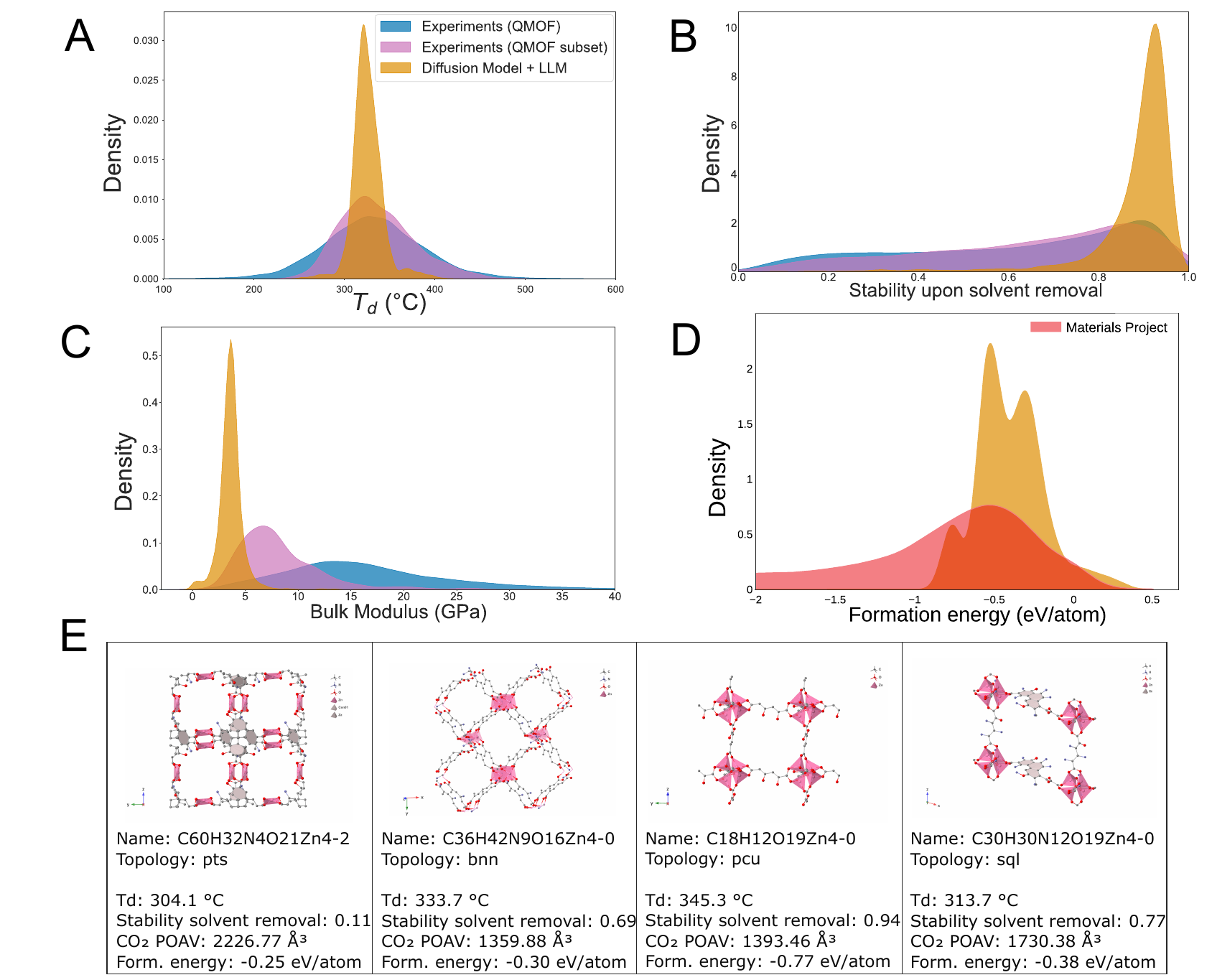}
    \caption{\textbf{Analysis of MOF crystal structures selected by \emph{QHarden}.} 
    \textbf{a}, Predicted decomposition temperature distribution ($T_{d}$),
    \textbf{b}, stability upon solvent removal \cite{nandy_using_2021},
    \textbf{c} and MACE-MP-0b computed Bulk modulus distributions for the Experimental MOF structures (QMOF database \cite{rosen_machine_2021}) (blue), the QMOF subset with chemical compositions matching diffusion-generated MOFs as well as pcu topology (pink), and the diffusion-generated MOFs, denoted Diffusion Model + LLM (orange).
    \textbf{d}, Formation energy computed at the r$^2$SCAN-D4 level for diffusion-generated MOFs compared to the Material Project data. 
    \textbf{e}, Representative samples of the ``AI-dreamt'' MOFs as well as their corresponding topologies and computed properties (Td: decomposition temperature; POAV: probe-occupiable accessible volume).}
    \label{fig:figure3}
\end{figure}

\begin{figure}
    \centering
    \includegraphics[width=0.9\textwidth]{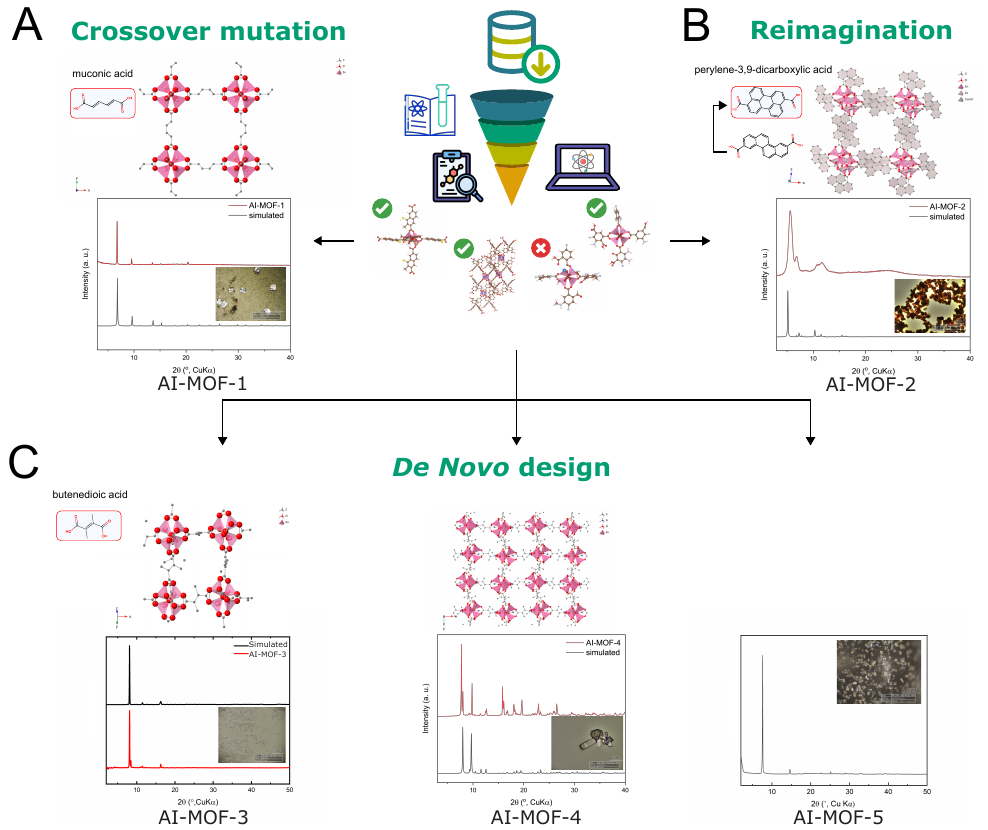}
    \caption{\textbf{Synthesis strategies and crystal structures of MOFs.} 
    Overview of the different strategies used to synthesize the MOFs, alongside the crystal structures, PXRD patterns with experimental (red) and simulated (black), and organic linkers. After filtering the crystal structures with \emph{QHarden}, the organic linkers of top MOF candidates were extracted and synthesized. 
    \textbf{a}, Crystal structure of AI-MOF-1, with corresponding organic linkers and experimental versus simulated PXRD patterns following the crossover mutation strategy. 
    \textbf{b}, Crystal structure of AI-MOF-2, showing the organic linkers before and after modification, with experimental and simulated PXRD patterns obtained from the reimagination strategy. 
    \textbf{c}, Crystal structures of AI-MOF-3, AI-MOF-4, and AI-MOF-5, PXRD patterns generated via the \textit{de novo} design strategy.}
    \label{fig:figure4}
\end{figure}

The successful synthesis of the MOF containing the deprotonated form of 2,3-dimethyl-2-butenedioic acid under two different experimental conditions correlates with the frequent generation of this linker by the model. Furthermore, each on these experimentally validated crystal obtained closely matched the diffusion-generated structure with MOF-5 topology. It is worth to note that the previous AI-MOFs exhibit MOF-5 topologies due to the filtering rules on reported zinc SBUs. However, as shown in Fig.\ref{fig:figure3}e, MOFGen is able to generate a variety of topology that could be used for future experiments.

To illustrate the versatility of diffusion-generated linkers with different metasl, we synthesized an aluminum-based MOF (AI-MOF-5) using 2,3-dimethyl-2-butenedioic acid, a linker frequently proposed by the model. While the PXRD suggest a MOF we weren't able to simulate its topology. The combination of aluminum-metal SBUs with hydrophobic dimethyl functional groups on the linker may allow the adjustment of the position of the onset in water sorption isotherms, broadening the potential use of AI-MOFs in water harvesting \cite{zheng_chatgpt_2023, zheng_shaping_2023}.

\section*{Discussion}

Our system of Agentic AI, MOFGen, is shown to accelerate the discovery of MOFs by integrating large-scale diffusion models with LLMs, quantum mechanical methods and synthesis predictor. MOFGen has greatly expanded the chemical space of MOFs, identifying 80,000 previously unreported, synthesizable organic linkers not present in the training. Further enhance by the combinatorial possibility through mixed-linker MOFs, broaden the chemical diversity accessible to future synthesis efforts. The robustness of our model's out-of-distribution generalization was experimentally validated by successfully synthesizing five ``AI-dreamt'' MOFs, marking the largest demonstration of synthesized diffusion-generated materials to date. Notably, all synthesized MOFs exhibited crystal structures and topologies in agreement with those generated by the diffusion model, significantly reducing the discovery timeline from years to weeks and enabling exploration of various applications ranging. 

Future efforts could expand this system of agentic AI to generate MOFs tailored with specific properties. Ultimately, this approach represents a significant advancement towards fully autonomous materials discovery laboratories. This approach offers a powerful new strategy for accelerating the discovery and synthesis of porous materials critical to addressing global challenges such as hydrogen storage, water harvesting, and chemical sensing. Our experimental validations highlight the potential of generative AI methodologies to reshape chemistry through application to large and complex material classes such as MOFs while emphasizing the importance of maintaining human-in-the-loop to effectively guide the exploration of synthesizable material spaces alongside AI-driven models.



\clearpage 

%
\bibliography{science_template} 

\begin{thebibliography}{100}
\providecommand{\url}[1]{\texttt{#1}}
\expandafter\ifx\csname urlstyle\endcsname\relax
  \providecommand{\doi}[1]{doi:\discretionary{}{}{}#1}\else
  \providecommand{\doi}{doi:\discretionary{}{}{}\begingroup \urlstyle{rm}\Url}\fi

\bibitem{lin_scalable_2021}
J.-B. Lin, \emph{et~al.}, A scalable metal-organic framework as a durable physisorbent for carbon dioxide capture \textbf{374}~(6574), 1464--1469, publisher: American Association for the Advancement of Science, \doi{10.1126/science.abi7281}, \url{https://www.science.org/doi/10.1126/science.abi7281}.

\bibitem{zhou_carbon_2024}
Z.~Zhou, \emph{et~al.}, Carbon dioxide capture from open air using covalent organic frameworks \textbf{635}~(8037), 96--101, publisher: Nature Publishing Group, \doi{10.1038/s41586-024-08080-x}, \url{https://www.nature.com/articles/s41586-024-08080-x}.

\bibitem{rohde_high-temperature_2024}
R.~C. Rohde, \emph{et~al.}, High-temperature carbon dioxide capture in a porous material with terminal zinc hydride sites \textbf{386}~(6723), 814--819, \doi{10.1126/science.adk5697}, \url{https://www.science.org/doi/10.1126/science.adk5697}.

\bibitem{rieth_record_2017}
A.~J. Rieth, S.~Yang, E.~N. Wang, M.~Dincă, Record Atmospheric Fresh Water Capture and Heat Transfer with a Material Operating at the Water Uptake Reversibility Limit \textbf{3}~(6), 668--672, \doi{10.1021/acscentsci.7b00186}, \url{https://doi.org/10.1021/acscentsci.7b00186}.

\bibitem{towsif_abtab_reticular_2018}
S.~M. Towsif~Abtab, \emph{et~al.}, Reticular Chemistry in Action: A Hydrolytically Stable {MOF} Capturing Twice Its Weight in Adsorbed Water \textbf{4}~(1), 94--105, \doi{10.1016/j.chempr.2017.11.005}, \url{https://www.sciencedirect.com/science/article/pii/S2451929417304734}.

\bibitem{hanikel_evolution_2021}
N.~Hanikel, \emph{et~al.}, Evolution of water structures in metal-organic frameworks for improved atmospheric water harvesting \textbf{374}~(6566), 454--459, publisher: American Association for the Advancement of Science, \doi{10.1126/science.abj0890}, \url{https://www.science.org/doi/full/10.1126/science.abj0890}.

\bibitem{zheng_high-yield_2023}
Z.~Zheng, \emph{et~al.}, High-yield, green and scalable methods for producing {MOF}-303 for water harvesting from desert air \textbf{18}~(1), 136--156, publisher: Nature Publishing Group, \doi{10.1038/s41596-022-00756-w}, \url{https://www.nature.com/articles/s41596-022-00756-w}.

\bibitem{hohenberg_inhomogeneous_1964}
P.~Hohenberg, W.~Kohn, Inhomogeneous Electron Gas \textbf{136}~(3), B864--B871, publisher: American Physical Society, \doi{10.1103/PhysRev.136.B864}, \url{https://link.aps.org/doi/10.1103/PhysRev.136.B864}.

\bibitem{kohn_self-consistent_1965}
W.~Kohn, L.~J. Sham, Self-Consistent Equations Including Exchange and Correlation Effects \textbf{140}~(4), A1133--A1138, publisher: American Physical Society, \doi{10.1103/PhysRev.140.A1133}, \url{https://link.aps.org/doi/10.1103/PhysRev.140.A1133}.

\bibitem{jain_commentary_2013}
A.~Jain, \emph{et~al.}, Commentary: The Materials Project: A materials genome approach to accelerating materials innovation \textbf{1}~(1), 011002, \doi{10.1063/1.4812323}, \url{https://doi.org/10.1063/1.4812323}.

\bibitem{ghahremanpour_alexandria_2018}
M.~M. Ghahremanpour, P.~J. van Maaren, D.~van~der Spoel, The Alexandria library, a quantum-chemical database of molecular properties for force field development \textbf{5}~(1), 180062, \doi{10.1038/sdata.2018.62}, \url{https://www.nature.com/articles/sdata201862}.

\bibitem{sriram_open_2024}
A.~Sriram, \emph{et~al.}, The Open {DAC} 2023 Dataset and Challenges for Sorbent Discovery in Direct Air Capture \textbf{10}~(5), 923--941, \doi{10.1021/acscentsci.3c01629}, \url{https://doi.org/10.1021/acscentsci.3c01629}.

\bibitem{rosen_machine_2021}
A.~S. Rosen, \emph{et~al.}, Machine learning the quantum-chemical properties of metal–organic frameworks for accelerated materials discovery \textbf{4}~(5), 1578--1597, \doi{10.1016/j.matt.2021.02.015}, \url{https://www.sciencedirect.com/science/article/pii/S2590238521000709}.

\bibitem{chung_advances_2019}
Y.~G. Chung, \emph{et~al.}, Advances, Updates, and Analytics for the Computation-Ready, Experimental Metal–Organic Framework Database: {CoRE} {MOF} 2019 \textbf{64}~(12), 5985--5998, \doi{10.1021/acs.jced.9b00835}, \url{https://doi.org/10.1021/acs.jced.9b00835}.

\bibitem{zhao_core_2024}
G.~Zhao, \emph{et~al.}, {CoRE} {MOF} {DB}: a curated experimental metal-organic framework database with machine-learned properties for integrated material-process screening, \doi{10.26434/chemrxiv-2024-nvmnr}, \url{https://chemrxiv.org/engage/chemrxiv/article-details/6757ca12f9980725cf91c7e0}.

\bibitem{szymanski_autonomous_2023}
N.~J. Szymanski, \emph{et~al.}, An autonomous laboratory for the accelerated synthesis of novel materials \textbf{624}~(7990), 86--91, \doi{10.1038/s41586-023-06734-w}, \url{https://www.nature.com/articles/s41586-023-06734-w}.

\bibitem{rao_machine_2022}
Z.~Rao, \emph{et~al.}, Machine learning–enabled high-entropy alloy discovery \textbf{378}~(6615), 78--85, \doi{10.1126/science.abo4940}, \url{https://www.science.org/doi/10.1126/science.abo4940}.

\bibitem{domingues_using_2022}
N.~P. Domingues, \emph{et~al.}, Using genetic algorithms to systematically improve the synthesis conditions of Al-{PMOF} \textbf{5}~(1), 1--8, \doi{10.1038/s42004-022-00785-2}, \url{https://www.nature.com/articles/s42004-022-00785-2}.

\bibitem{zeni_generative_2025}
C.~Zeni, \emph{et~al.}, A generative model for inorganic materials design pp. 1--3, \doi{10.1038/s41586-025-08628-5}, \url{https://www.nature.com/articles/s41586-025-08628-5}.

\bibitem{yang_scalable_2024}
S.~Yang, \emph{et~al.}, Scalable Diffusion for Materials Generation, \doi{10.48550/arXiv.2311.09235}, \url{http://arxiv.org/abs/2311.09235}.

\bibitem{fu_mofdiff_2023}
X.~Fu, T.~Xie, A.~S. Rosen, T.~Jaakkola, J.~Smith, {MOFDiff}: Coarse-grained Diffusion for Metal-Organic Framework Design, \doi{10.48550/arXiv.2310.10732}, \url{http://arxiv.org/abs/2310.10732}.

\bibitem{behler_generalized_2007}
J.~Behler, M.~Parrinello, Generalized Neural-Network Representation of High-Dimensional Potential-Energy Surfaces \textbf{98}~(14), 146401, \doi{10.1103/PhysRevLett.98.146401}, \url{https://link.aps.org/doi/10.1103/PhysRevLett.98.146401}.

\bibitem{smith_ani-1_2017}
J.~S. Smith, O.~Isayev, A.~E. Roitberg, {ANI}-1: an extensible neural network potential with {DFT} accuracy at force field computational cost \textbf{8}~(4), 3192--3203, \doi{10.1039/C6SC05720A}, \url{https://pubs.rsc.org/en/content/articlelanding/2017/sc/c6sc05720a}.

\bibitem{schutt_schnet_2018}
K.~T. Schütt, H.~E. Sauceda, P.-J. Kindermans, A.~Tkatchenko, K.-R. Müller, {SchNet} – A deep learning architecture for molecules and materials \textbf{148}~(24), 241722, \doi{10.1063/1.5019779}, \url{https://doi.org/10.1063/1.5019779}.

\bibitem{wang_deepmd-kit_2018}
H.~Wang, L.~Zhang, J.~Han, W.~E, {DeePMD}-kit: A deep learning package for many-body potential energy representation and molecular dynamics \textbf{228}, 178--184, \doi{10.1016/j.cpc.2018.03.016}, \url{https://www.sciencedirect.com/science/article/pii/S0010465518300882}.

\bibitem{batzner_e3-equivariant_2022}
S.~Batzner, \emph{et~al.}, E(3)-equivariant graph neural networks for data-efficient and accurate interatomic potentials \textbf{13}~(1), 2453, \doi{10.1038/s41467-022-29939-5}, \url{https://www.nature.com/articles/s41467-022-29939-5}.

\bibitem{inizan_scalable_2023}
T.~J. Inizan, \emph{et~al.}, Scalable hybrid deep neural networks/polarizable potentials biomolecular simulations including long-range effects \textbf{14}~(20), 5438--5452, \doi{10.1039/D2SC04815A}, \url{https://pubs.rsc.org/en/content/articlelanding/2023/sc/d2sc04815a}.

\bibitem{ple_force-field-enhanced_2023}
T.~Plé, L.~Lagardère, J.-P. Piquemal, Force-field-enhanced neural network interactions: from local equivariant embedding to atom-in-molecule properties and long-range effects \textbf{14}~(44), 12554--12569, \doi{10.1039/D3SC02581K}, \url{https://pubs.rsc.org/en/content/articlelanding/2023/sc/d3sc02581k}.

\bibitem{batatia_mace_2023}
I.~Batatia, D.~P. Kovács, G.~N.~C. Simm, C.~Ortner, G.~Csányi, {MACE}: Higher Order Equivariant Message Passing Neural Networks for Fast and Accurate Force Fields, \doi{10.48550/arXiv.2206.07697}, \url{http://arxiv.org/abs/2206.07697}.

\bibitem{batatia_foundation_2024}
I.~Batatia, \emph{et~al.}, A foundation model for atomistic materials chemistry, \doi{10.48550/arXiv.2401.00096}, \url{http://arxiv.org/abs/2401.00096}.

\bibitem{merchant_scaling_2023}
A.~Merchant, \emph{et~al.}, Scaling deep learning for materials discovery \textbf{624}~(7990), 80--85, \doi{10.1038/s41586-023-06735-9}, \url{https://www.nature.com/articles/s41586-023-06735-9}.

\bibitem{yaghi_reticular_2003}
O.~M. Yaghi, \emph{et~al.}, Reticular synthesis and the design of new materials \textbf{423}~(6941), 705--714, number: 6941 Publisher: Nature Publishing Group, \doi{10.1038/nature01650}, \url{https://www.nature.com/articles/nature01650}.

\bibitem{furukawa_chemistry_2013}
H.~Furukawa, K.~E. Cordova, M.~O’Keeffe, O.~M. Yaghi, The Chemistry and Applications of Metal-Organic Frameworks \textbf{341}~(6149), 1230444, publisher: American Association for the Advancement of Science, \doi{10.1126/science.1230444}, \url{https://www.science.org/doi/abs/10.1126/science.1230444}.

\bibitem{zheng_large_2025}
Z.~Zheng, \emph{et~al.}, Large language models for reticular chemistry pp. 1--13, publisher: Nature Publishing Group, \doi{10.1038/s41578-025-00772-8}, \url{https://www.nature.com/articles/s41578-025-00772-8}.

\bibitem{park_generative_2024}
H.~Park, \emph{et~al.}, A generative artificial intelligence framework based on a molecular diffusion model for the design of metal-organic frameworks for carbon capture \textbf{7}~(1), 1--18, \doi{10.1038/s42004-023-01090-2}, \url{https://www.nature.com/articles/s42004-023-01090-2}.

\bibitem{jumper_highly_2021}
J.~Jumper, \emph{et~al.}, Highly accurate protein structure prediction with {AlphaFold} \textbf{596}~(7873), 583--589, \doi{10.1038/s41586-021-03819-2}, \url{https://www.nature.com/articles/s41586-021-03819-2}.

\bibitem{abramson_accurate_2024}
J.~Abramson, \emph{et~al.}, Accurate structure prediction of biomolecular interactions with {AlphaFold} 3 \textbf{630}~(8016), 493--500, \doi{10.1038/s41586-024-07487-w}, \url{https://www.nature.com/articles/s41586-024-07487-w}.

\bibitem{yang_generative_2024}
S.~Yang, \emph{et~al.}, Generative Hierarchical Materials Search, \doi{10.48550/arXiv.2409.06762}, \url{http://arxiv.org/abs/2409.06762}.

\bibitem{bartel_review_2022}
C.~J. Bartel, Review of computational approaches to predict the thermodynamic stability of inorganic solids \textbf{57}~(23), 10475--10498, \doi{10.1007/s10853-022-06915-4}, \url{https://doi.org/10.1007/s10853-022-06915-4}.

\bibitem{openai_gpt-4_2024}
{OpenAI}, \emph{et~al.}, {GPT}-4 Technical Report, \doi{10.48550/arXiv.2303.08774}, \url{http://arxiv.org/abs/2303.08774}.

\bibitem{touvron_llama_2023}
H.~Touvron, \emph{et~al.}, {LLaMA}: Open and Efficient Foundation Language Models, \doi{10.48550/arXiv.2302.13971}, \url{http://arxiv.org/abs/2302.13971}.

\bibitem{sohl-dickstein_deep_2015}
J.~Sohl-Dickstein, E.~A. Weiss, N.~Maheswaranathan, S.~Ganguli, Deep Unsupervised Learning using Nonequilibrium Thermodynamics, \doi{10.48550/arXiv.1503.03585}, \url{http://arxiv.org/abs/1503.03585}.

\bibitem{ho_denoising_2020}
J.~Ho, A.~Jain, P.~Abbeel, Denoising Diffusion Probabilistic Models, \doi{10.48550/arXiv.2006.11239}, \url{http://arxiv.org/abs/2006.11239}.

\bibitem{perdew_generalized_1996}
J.~P. Perdew, K.~Burke, M.~Ernzerhof, Generalized Gradient Approximation Made Simple \textbf{77}~(18), 3865--3868, publisher: American Physical Society, \doi{10.1103/PhysRevLett.77.3865}, \url{https://link.aps.org/doi/10.1103/PhysRevLett.77.3865}.

\bibitem{furness_accurate_2020}
J.~W. Furness, A.~D. Kaplan, J.~Ning, J.~P. Perdew, J.~Sun, Accurate and Numerically Efficient r2SCAN Meta-Generalized Gradient Approximation \textbf{11}~(19), 8208--8215, \doi{10.1021/acs.jpclett.0c02405}, \url{https://doi.org/10.1021/acs.jpclett.0c02405}.

\bibitem{coley_scscore_2018}
C.~W. Coley, L.~Rogers, W.~H. Green, K.~F. Jensen, {SCScore}: Synthetic Complexity Learned from a Reaction Corpus \textbf{58}~(2), 252--261, publisher: American Chemical Society, \doi{10.1021/acs.jcim.7b00622}, \url{https://doi.org/10.1021/acs.jcim.7b00622}.

\bibitem{nandy_using_2021}
A.~Nandy, C.~Duan, H.~J. Kulik, Using Machine Learning and Data Mining to Leverage Community Knowledge for the Engineering of Stable Metal–Organic Frameworks \textbf{143}~(42), 17535--17547, publisher: American Chemical Society, \doi{10.1021/jacs.1c07217}, \url{https://doi.org/10.1021/jacs.1c07217}.

\bibitem{wei_chain--thought_2023}
J.~Wei, \emph{et~al.}, Chain-of-Thought Prompting Elicits Reasoning in Large Language Models, \doi{10.48550/arXiv.2201.11903}, \url{http://arxiv.org/abs/2201.11903}.

\bibitem{noauthor_largescale_nodate}
Large‐Scale Production of Metal–Organic Frameworks - Chakraborty - 2024 - Advanced Functional Materials - Wiley Online Library, \url{https://advanced.onlinelibrary.wiley.com/doi/full/10.1002%2Fadfm.202309089}.

\bibitem{paul_scale-up_2023}
T.~Paul, \emph{et~al.}, Scale-up of metal-organic frameworks production: Engineering strategies and prospects towards sustainable manufacturing \textbf{11}~(5), 111112, \doi{10.1016/j.jece.2023.111112}, \url{https://www.sciencedirect.com/science/article/pii/S2213343723018511}.

\bibitem{willems_algorithms_2012}
T.~F. Willems, C.~H. Rycroft, M.~Kazi, J.~C. Meza, M.~Haranczyk, Algorithms and tools for high-throughput geometry-based analysis of crystalline porous materials \textbf{149}~(1), 134--141, \doi{10.1016/j.micromeso.2011.08.020}, \url{https://www.sciencedirect.com/science/article/pii/S1387181111003738}.

\bibitem{elena_machine_2024}
A.~M. Elena, \emph{et~al.}, Machine Learned Potential for High-Throughput Phonon Calculations of Metal-Organic Frameworks, \doi{10.48550/arXiv.2412.02877}, \url{http://arxiv.org/abs/2412.02877}.

\bibitem{lim_accelerating_2024}
Y.~Lim, H.~Park, A.~Walsh, J.~Kim, Accelerating {CO}2 Direct Air Capture Screening for Metal-Organic Frameworks with a Transferable Machine Learning Force Field, \doi{10.26434/chemrxiv-2024-7w6g6}, \url{https://chemrxiv.org/engage/chemrxiv/article-details/6759b06df9980725cfbc8cef}.

\bibitem{wolos_computer-designed_2022}
A.~Wołos, \emph{et~al.}, Computer-designed repurposing of chemical wastes into drugs \textbf{604}~(7907), 668--676, publisher: Nature Publishing Group, \doi{10.1038/s41586-022-04503-9}, \url{https://www.nature.com/articles/s41586-022-04503-9}.

\bibitem{zadlo-dobrowolska_computational_2024}
A.~zadłl Dobrowolska, \emph{et~al.}, Computational synthesis design for controlled degradation and revalorization \textbf{3}~(5), 643--654, publisher: Nature Publishing Group, \doi{10.1038/s44160-024-00497-6}, \url{https://www.nature.com/articles/s44160-024-00497-6}.

\bibitem{strieth-kalthoff_artificial_2024}
F.~Strieth-Kalthoff, \emph{et~al.}, Artificial Intelligence for Retrosynthetic Planning Needs Both Data and Expert Knowledge \textbf{146}~(16), 11005--11017, publisher: American Chemical Society, \doi{10.1021/jacs.4c00338}, \url{https://doi.org/10.1021/jacs.4c00338}.

\bibitem{groom_cambridge_2016}
C.~R. Groom, I.~J. Bruno, M.~P. Lightfoot, S.~C. Ward, The Cambridge Structural Database \textbf{72}~(2), 171--179, \doi{10.1107/S2052520616003954}, \url{//journals.iucr.org/paper?bm5086}.

\bibitem{chen_estimating_2024}
S.~Chen, Y.~Jung, Estimating the synthetic accessibility of molecules with building block and reaction-aware {SAScore} \textbf{16}~(1), 83, \doi{10.1186/s13321-024-00879-0}, \url{https://doi.org/10.1186/s13321-024-00879-0}.

\bibitem{barthelet_breathing_2002}
K.~Barthelet, J.~Marrot, D.~Riou, G.~Férey, A Breathing Hybrid Organic–Inorganic Solid with Very Large Pores and High Magnetic Characteristics \textbf{41}~(2), 281--284, \_eprint: https://onlinelibrary.wiley.com/doi/pdf/10.1002/1521-3773\%2820020118\%2941\%3A2\%3C281\%3A\%3AAID-{ANIE}281\%3E3.0.{CO}\%3B2-Y, \doi{10.1002/1521-3773(20020118)41:2<281::AID-ANIE281>3.0.CO;2-Y}, \url{https://onlinelibrary.wiley.com/doi/abs/10.1002/1521-3773%2820020118%2941%3A2%3C281%3A%3AAID-ANIE281%3E3.0.CO%3B2-Y}.

\bibitem{loiseau_rationale_2004}
T.~Loiseau, \emph{et~al.}, A rationale for the large breathing of the porous aluminum terephthalate ({MIL}-53) upon hydration \textbf{10}~(6), 1373--1382, \doi{10.1002/chem.200305413}.

\bibitem{serre_explanation_2007}
C.~Serre, \emph{et~al.}, An Explanation for the Very Large Breathing Effect of a Metal–Organic Framework during {CO}$_{\textrm{2}}$ Adsorption \textbf{19}~(17), 2246--2251, \doi{10.1002/adma.200602645}, \url{https://onlinelibrary.wiley.com/doi/10.1002/adma.200602645}.

\bibitem{alhamami_review_2014}
M.~Alhamami, H.~Doan, C.-H. Cheng, A Review on Breathing Behaviors of Metal-Organic-Frameworks ({MOFs}) for Gas Adsorption \textbf{7}~(4), 3198--3250, \doi{10.3390/ma7043198}, \url{https://www.ncbi.nlm.nih.gov/pmc/articles/PMC5453333/}.

\bibitem{ntep_designing_2019}
T.~J. M.~M. Ntep, \emph{et~al.}, Designing a new aluminium muconate metal–organic framework ({MIL}-53-muc) as a methanol adsorbent for sub-zero temperature heat transformation applications \textbf{7}~(43), 24973--24981, publisher: The Royal Society of Chemistry, \doi{10.1039/C9TA07465A}, \url{https://pubs.rsc.org/en/content/articlelanding/2019/ta/c9ta07465a}.

\bibitem{mof5_2005}
J.~L.~C. Rowsell, E.~C. Spencer, J.~Eckert, J.~A.~K. Howard, O.~M. Yaghi, Gas Adsorption Sites in a Large-Pore Metal-Organic Framework. \emph{Science} \textbf{309}~(5739), 1350--1354 (2005), \doi{10.1126/science.1113247}, \url{https://www.science.org/doi/abs/10.1126/science.1113247}.

\bibitem{qiaowei_2009}
Q.~Li, \emph{et~al.}, Docking in Metal-Organic Frameworks. \emph{Science} \textbf{325}~(5942), 855--859 (2009), \doi{10.1126/science.1175441}, \url{https://www.science.org/doi/abs/10.1126/science.1175441}.

\bibitem{brozek_dynamic_2015}
C.~K. Brozek, \emph{et~al.}, Dynamic {DMF} Binding in {MOF}-5 Enables the Formation of Metastable Cobalt-Substituted {MOF}-5 Analogues \textbf{1}~(5), 252--260, publisher: American Chemical Society, \doi{10.1021/acscentsci.5b00247}, \url{https://doi.org/10.1021/acscentsci.5b00247}.

\bibitem{dzubak_ab_2012}
A.~L. Dzubak, \emph{et~al.}, Ab initio carbon capture in open-site metal–organic frameworks \textbf{4}~(10), 810--816, \doi{10.1038/nchem.1432}, \url{https://www.nature.com/articles/nchem.1432}.

\bibitem{wei_metalorganic_2020}
Y.-S. Wei, M.~Zhang, R.~Zou, Q.~Xu, Metal–Organic Framework-Based Catalysts with Single Metal Sites \textbf{120}~(21), 12089--12174, publisher: American Chemical Society, \doi{10.1021/acs.chemrev.9b00757}, \url{https://doi.org/10.1021/acs.chemrev.9b00757}.

\bibitem{zheng_chatgpt_2023}
Z.~Zheng, O.~Zhang, C.~Borgs, J.~T. Chayes, O.~M. Yaghi, {ChatGPT} Chemistry Assistant for Text Mining and the Prediction of {MOF} Synthesis \textbf{145}~(32), 18048--18062, \doi{10.1021/jacs.3c05819}, \url{https://doi.org/10.1021/jacs.3c05819}.

\bibitem{zheng_shaping_2023}
Z.~Zheng, \emph{et~al.}, Shaping the Water-Harvesting Behavior of Metal–Organic Frameworks Aided by Fine-Tuned {GPT} Models \textbf{145}~(51), 28284--28295, publisher: American Chemical Society, \doi{10.1021/jacs.3c12086}, \url{https://doi.org/10.1021/jacs.3c12086}.

\bibitem{bucior_identification_2019}
B.~J. Bucior, \emph{et~al.}, Identification Schemes for Metal–Organic Frameworks To Enable Rapid Search and Cheminformatics Analysis \textbf{19}~(11), 6682--6697, publisher: American Chemical Society, \doi{10.1021/acs.cgd.9b01050}, \url{https://doi.org/10.1021/acs.cgd.9b01050}.

\bibitem{maaten_visualizing_2008}
L.~v.~d. Maaten, G.~Hinton, Visualizing Data using t-{SNE} \textbf{9}~(86), 2579--2605, \url{http://jmlr.org/papers/v9/vandermaaten08a.html}.

\bibitem{ertl_estimation_2009}
P.~Ertl, A.~Schuffenhauer, Estimation of synthetic accessibility score of drug-like molecules based on molecular complexity and fragment contributions \textbf{1}~(1), 8, \doi{10.1186/1758-2946-1-8}, \url{https://doi.org/10.1186/1758-2946-1-8}.

\bibitem{dolomanov_olex2_2009}
O.~V. Dolomanov, L.~J. Bourhis, R.~J. Gildea, J.~a.~K. Howard, H.~Puschmann, {OLEX}2: a complete structure solution, refinement and analysis program \textbf{42}~(2), 339--341, publisher: International Union of Crystallography, \doi{10.1107/S0021889808042726}, \url{//journals.iucr.org/paper?kk5042}.

\bibitem{bruno_new_2002}
I.~J. Bruno, \emph{et~al.}, New software for searching the Cambridge Structural Database and visualizing crystal structures \textbf{58}~(3), 389--397, publisher: International Union of Crystallography, \doi{10.1107/S0108768102003324}, \url{https://journals.iucr.org/b/issues/2002/03/01/an0609/}.

\bibitem{sykes_what_2024}
R.~A. Sykes, \emph{et~al.}, What has scripting ever done for us? The {CSD} Python application programming interface ({API}) \textbf{57}~(4), 1235--1250, publisher: International Union of Crystallography, \doi{10.1107/S1600576724005934}, \url{https://journals.iucr.org/j/issues/2024/04/00/oc5038/}.

\bibitem{nandy_database_2023}
A.~Nandy, \emph{et~al.}, A database of ultrastable {MOFs} reassembled from stable fragments with machine learning models \textbf{6}~(5), 1585--1603, \doi{10.1016/j.matt.2023.03.009}, \url{https://www.sciencedirect.com/science/article/pii/S259023852300111X}.

\bibitem{boyd_data-driven_2019}
P.~G. Boyd, \emph{et~al.}, Data-driven design of metal–organic frameworks for wet flue gas {CO}2 capture \textbf{576}~(7786), 253--256, publisher: Nature Publishing Group, \doi{10.1038/s41586-019-1798-7}, \url{https://www.nature.com/articles/s41586-019-1798-7}.

\bibitem{gomez-gualdron_evaluating_2016}
D.~A. Gómez-Gualdrón, \emph{et~al.}, Evaluating topologically diverse metal–organic frameworks for cryo-adsorbed hydrogen storage \textbf{9}~(10), 3279--3289, publisher: The Royal Society of Chemistry, \doi{10.1039/C6EE02104B}, \url{https://pubs.rsc.org/en/content/articlelanding/2016/ee/c6ee02104b}.

\bibitem{ganose2025atomate2}
A.~Ganose, \emph{et~al.}, Atomate2: Modular workflows for materials science (2025), \doi{10.26434/chemrxiv-2025-tcr5h}, \url{http://dx.doi.org/10.26434/chemrxiv-2025-tcr5h}.

\bibitem{grimme_robust_2017}
S.~Grimme, C.~Bannwarth, P.~Shushkov, A Robust and Accurate Tight-Binding Quantum Chemical Method for Structures, Vibrational Frequencies, and Noncovalent Interactions of Large Molecular Systems Parametrized for All spd-Block Elements (Z = 1–86) \textbf{13}~(5), 1989--2009, \doi{10.1021/acs.jctc.7b00118}, \url{https://doi.org/10.1021/acs.jctc.7b00118}.

\bibitem{larsen2017ase}
A.~H. Larsen, \emph{et~al.}, The atomic simulation environment—a Python library for working with atoms. \emph{J. Phys. Condens. Matter} \textbf{29}~(27), 273002 (2017), \doi{10.1088/1361-648X/aa680e}.

\bibitem{kresse_ab_1993}
G.~Kresse, J.~Hafner, Ab initio molecular dynamics for liquid metals \textbf{47}~(1), 558--561, publisher: American Physical Society, \doi{10.1103/PhysRevB.47.558}, \url{https://link.aps.org/doi/10.1103/PhysRevB.47.558}.

\bibitem{kresse1994vasp2}
G.~Kresse, J.~Hafner, \textit{{Ab} initio} molecular-dynamics simulation of the liquid-metal–amorphous-semiconductor transition in germanium. \emph{Physical Review B} \textbf{49}~(20), 14251--14269 (1994), \doi{10.1103/PhysRevB.49.14251}, \url{https://link.aps.org/doi/10.1103/PhysRevB.49.14251}.

\bibitem{kresse_efficiency_1996}
G.~Kresse, J.~Furthmüller, Efficiency of ab-initio total energy calculations for metals and semiconductors using a plane-wave basis set \textbf{6}~(1), 15--50, \doi{10.1016/0927-0256(96)00008-0}, \url{https://www.sciencedirect.com/science/article/pii/0927025696000080}.

\bibitem{kresse_efficient_1996}
G.~Kresse, J.~Furthmüller, Efficient iterative schemes for ab initio total-energy calculations using a plane-wave basis set \textbf{54}~(16), 11169--11186, publisher: American Physical Society, \doi{10.1103/PhysRevB.54.11169}, \url{https://link.aps.org/doi/10.1103/PhysRevB.54.11169}.

\bibitem{caldeweyher2019d4}
E.~Caldeweyher, \emph{et~al.}, A generally applicable atomic-charge dependent London dispersion correction. \emph{J. Chem. Phys.} \textbf{150}~(15), 154122 (2019), \doi{10.1063/1.5090222}, \url{https://doi.org/10.1063/1.5090222}.

\bibitem{ehlert2021r2scand4}
S.~Ehlert, \emph{et~al.}, r2SCAN-D4: Dispersion corrected meta-generalized gradient approximation for general chemical applications. \emph{J. Chem. Phys.} \textbf{154}~(6), 061101 (2021), \doi{10.1063/5.0041008}, \url{https://doi.org/10.1063/5.0041008}.

\bibitem{ong_python_2013}
S.~P. Ong, \emph{et~al.}, Python Materials Genomics (pymatgen): A robust, open-source python library for materials analysis \textbf{68}, 314--319, \doi{10.1016/j.commatsci.2012.10.028}, \url{https://www.sciencedirect.com/science/article/pii/S0927025612006295}.

\bibitem{barrosoluque2024omat}
L.~Barroso-Luque, \emph{et~al.}, Open Materials 2024 (OMat24) Inorganic Materials Dataset and Models (2024), https://arXiv:2410.12771.

\bibitem{flam-shepherd_language_2022}
D.~Flam-Shepherd, K.~Zhu, A.~Aspuru-Guzik, Language models can learn complex molecular distributions \textbf{13}~(1), 3293, publisher: Nature Publishing Group, \doi{10.1038/s41467-022-30839-x}, \url{https://www.nature.com/articles/s41467-022-30839-x}.

\bibitem{m_bran_augmenting_2024}
A.~M.~Bran, \emph{et~al.}, Augmenting large language models with chemistry tools \textbf{6}~(5), 525--535, publisher: Nature Publishing Group, \doi{10.1038/s42256-024-00832-8}, \url{https://www.nature.com/articles/s42256-024-00832-8}.

\bibitem{cavanagh_smileyllama_2024}
J.~M. Cavanagh, \emph{et~al.}, {SmileyLlama}: Modifying Large Language Models for Directed Chemical Space Exploration, \doi{10.48550/arXiv.2409.02231}, \url{http://arxiv.org/abs/2409.02231}.

\bibitem{rosi_hydrogen_2003}
N.~L. Rosi, \emph{et~al.}, Hydrogen Storage in Microporous Metal-Organic Frameworks \textbf{300}~(5622), 1127--1129, publisher: American Association for the Advancement of Science, \doi{10.1126/science.1083440}, \url{https://www.science.org/doi/10.1126/science.1083440}.

\bibitem{bannwarth_gfn2-xtbaccurate_2019}
C.~Bannwarth, S.~Ehlert, S.~Grimme, {GFN}2-{xTB}—An Accurate and Broadly Parametrized Self-Consistent Tight-Binding Quantum Chemical Method with Multipole Electrostatics and Density-Dependent Dispersion Contributions \textbf{15}~(3), 1652--1671, publisher: American Chemical Society, \doi{10.1021/acs.jctc.8b01176}, \url{https://doi.org/10.1021/acs.jctc.8b01176}.

\bibitem{nurhuda_performance_2022}
M.~Nurhuda, C.~C. Perry, M.~A. Addicoat, Performance of {GFN}1-{xTB} for periodic optimization of metal organic frameworks \textbf{24}~(18), 10906--10914, publisher: Royal Society of Chemistry, \doi{10.1039/D2CP00184E}, \url{https://pubs.rsc.org/en/content/articlelanding/2022/cp/d2cp00184e}.

\bibitem{thi_le_evaluation_nodate}
T.-H. Thi~Le, P.~Gómez-Orellana, M.~A. Ortuño, Evaluation of Semiempirical Quantum Mechanical Methods for Zr-Based Metal–Organic Framework Catalysts \textbf{n/a}, e202400588, \_eprint: https://onlinelibrary.wiley.com/doi/pdf/10.1002/cphc.202400588, \doi{10.1002/cphc.202400588}, \url{https://onlinelibrary.wiley.com/doi/abs/10.1002/cphc.202400588}.

\bibitem{cavka_new_2008}
J.~H. Cavka, \emph{et~al.}, A New Zirconium Inorganic Building Brick Forming Metal Organic Frameworks with Exceptional Stability \textbf{130}~(42), 13850--13851, publisher: American Chemical Society, \doi{10.1021/ja8057953}, \url{https://doi.org/10.1021/ja8057953}.

\bibitem{nandy_mofsimplify_2022}
A.~Nandy, \emph{et~al.}, {MOFSimplify}, machine learning models with extracted stability data of three thousand metal–organic frameworks \textbf{9}~(1), 74, publisher: Nature Publishing Group, \doi{10.1038/s41597-022-01181-0}, \url{https://www.nature.com/articles/s41597-022-01181-0}.

\bibitem{healy_thermal_2020}
C.~Healy, \emph{et~al.}, The thermal stability of metal-organic frameworks \textbf{419}, 213388, \doi{10.1016/j.ccr.2020.213388}, \url{https://www.sciencedirect.com/science/article/pii/S0010854520302836}.

\bibitem{eckhoff_molecular_2019}
M.~Eckhoff, J.~Behler, From Molecular Fragments to the Bulk: Development of a Neural Network Potential for {MOF}-5 \textbf{15}~(6), 3793--3809, publisher: American Chemical Society, \doi{10.1021/acs.jctc.8b01288}, \url{https://doi.org/10.1021/acs.jctc.8b01288}.

\bibitem{wieser_machine_2024}
S.~Wieser, E.~Zojer, Machine learned force-fields for an Ab-initio quality description of metal-organic frameworks \textbf{10}~(1), 1--18, publisher: Nature Publishing Group, \doi{10.1038/s41524-024-01205-w}, \url{https://www.nature.com/articles/s41524-024-01205-w}.

\end{thebibliography}
\bibliographystyle{sciencemag}

%
%
%
%
%
%


\section*{Acknowledgments}
T.J.I. thanks Prof. Andrew Rosen for discussion about the initial workflow. T.J.I. thanks Dr. Saumil Chheda for discussion about the zeo++ criteria.
\paragraph*{Funding:}
This work was intellecturally led by the Bakar Institute of Digital Materials for the Planet (BIDMaP). Additional support on data generation and workflow development was obtained from the U.S. Department of Energy, Office of Science, Office of Basic Energy Sciences, Materials Sciences and Engineering Division under contract No. DE-AC02-05-CH11231 (Materials Project program KC23MP).
This research used resources of the National Energy Research Scientific Computing Center (NERSC), a Department of Energy Office of Science User Facility using NERSC award DOE-ERCAP0026371, DOE-ERCAP0032604, DOE-ERCAP0032988 and DOE-ERCAP0031751 'GenAI@NERSC'.
\paragraph*{Author contributions:}
T.J.I. conceptualized the project; T.J.I., A.K. developed and executed each stage of the quantum mechanical agents; T.J.I. and Y.L. curated and extracted the databases used for training; Y.L. fine-tuned the LLM on organic linkers and generated the chemical formulae; S.Y. trained the diffusion model. Y.L., J.Y., A.H.A. and S.M. conducted automated synthesis, XRD experiments and solved the crystal structures; R.C. assessed GFN1-xTB precision on MOFs; R.C. and T.J.I. performed and tested deep learning synthesizability predictors; R.C., A.K. and T.J.I. performed bulk modulus analysis using the MACE-MP-0 model; M.A. and T.J.I. assessed synthesizability using Allchemy software; M.A. analyzed the synthesizability and the chemistry of the generated organic linkers; Z.Z., M.A., Y.L., J.Y., A.H.A. and S.M. developed expert-designed rules for ranking organic linkers; M.A., A.K. and T.J.I. developed post-processing algorithms for crystal structures; T.J.I., O.M.Y., and K.P. prepared the initial draft; and all authors contributed to revising the manuscript.
\paragraph*{Competing interests:}
``There are no competing interests to declare.''
\paragraph*{Data and materials availability:}
The crystal structures are available from the Cambridge Crystallographic Data Center under the reference numbers 2433335 and 2442597, AI-MOF-4 and AI-MOF-7, respectively. The final database, DFT computations, and structure visualization are publicly available on the Material Project website: https://next-gen.materialsproject.org/contribs/projects/MOFGen\_2025


\subsection*{Supplementary materials}
Materials and Methods\\
Supplementary Text\\
Figures S1 to S19\\
Tables S1 to S3\\
References \textit{(7-\arabic{enumiv})}\\ 
Checkcif S1 to S2 


\newpage


\renewcommand{\thefigure}{S\arabic{figure}}
\renewcommand{\thetable}{S\arabic{table}}
\renewcommand{\theequation}{S\arabic{equation}}
\renewcommand{\thepage}{S\arabic{page}}
\setcounter{figure}{0}
\setcounter{table}{0}
\setcounter{equation}{0}
\setcounter{page}{1} 


\begin{center}
\section*{Supplementary Materials for\\ \scititle}

Théo~Jaffrelot~Inizan$^{\ast}$, Sherry~Yang$^{\dagger}$, Aaron~Kaplan$^{\dagger}$, Yen-hsu~Lin$^{\dagger}$,\\
Yen-hsu~Lin$^{\dagger}$, Jian~Yin, Saber~Mirzaei, Mona~Abdelgaid, Ali~H.~Alawadhi, \\
KwangHwan~Cho, Zhiling~Zheng, Ekin~Dogus~Cubuk, Christian~Borgs, \\
Jennifer~T.~Chayes, Kristin~A.~Persson$^{\ast}$, Omar~M.~Yaghi$^{\ast}$ \\
\small$^\ast$Corresponding author. Email: theo.jaf@berkeley.edu, kristinpersson@berkeley.edu, yaghi@berkeley.edu\\
\small$^\dagger$These authors contributed equally to this work.
\end{center}
\subsubsection*{This PDF file includes:}
Materials and Methods\\
Supplementary Text\\
Figures S1 to S19\\
Tables S1 to S3\\

\subsubsection*{Other Supplementary Materials for this manuscript:}
Checkcif S1 to S2

\newpage


\subsection{Materials and Methods}\label{SI:material_method_synth}

\subsubsection*{Chemicals}
Trans,trans-muconic acid, 2,3-dimethyl-2-butenedioic acid, sminomalonic acid, tartronic acid were purchased from Aaron Chemical LLC. 1,2,4-Triazole was purchased from TCI America. $N,N$-dimethylformamide (DMF) ,and $N,N$-diethylformamide (DEF) were purchased from AK Scientific. Zinc nitrate hexahydrate was purchased from Sigma-Aldrich. Perylene-3,9-dicarboxylic acid was purchased from ChemScene LLC. Unless specified, all chemical reagents were used without further purification. 

\subsubsection*{High-Throughput Synthesis}
OT-2 Liquid Handler was purchased from Opentrons Labworks Inc. OT-2 was controlled by Opentrons App and Python scripts. 96-well plates, 0.5 mL borosilicate glass containers, and polytetrafluoroethylene sheet were purchased from Analytical Sales and Services Inc. 

\subsubsection*{Microwave-Assisted Reaction}
Microwave reactions were conducted using the CEM Discover\textregistered Microwave Synthesizer.

\subsubsection*{Powder X-Ray Diffraction (PXRD)}
AI-MOF-2 and AI-MOF-4, PXRD patterns were collected using a Rigaku Miniflex 600 X-ray diffractometer in reflection geometry employing Cu K$\alpha$ radiation ($\lambda$ = 1.54184 \AA{}, with a Ni filter) at a power of 600 W (40 kV, 15 mA). Samples were mounted on Si (510) sample holders and leveled with a spatula. The step size was 0.02$^{\circ}$ with an exposure time of 0.5s per step. The other PXRD measurements were performed using the Bruker D8 Advance X-ray diffractometer equipped with a Cu anode and a Ni filter (Cu K$\alpha$$\textsubscript{1}$, $\lambda$ $=$ 1.54059 \AA{}) in Bragg-Brentano geometry. The samples were mounted on a zero-background holder and leveled using a glass slide.

\subsubsection*{Single-crystal X-ray Diffraction (SCXRD)}
SCXRD measurements were conducted on a Bruker Quest D8-Venture diffractometer equipped with a PHOTON100 CMOS detector and a micro-focus X-ray tube with a Cu target ($\lambda$ $=$ 1.54178 \AA{}. Single crystals were mounted on loops in oil and placed in nitrogen cold stream from Oxford Cryosystems equipment. The Bruker APEX4 software package was used for collecting, indexing, integrating, and scaling the data. The data were first integrated using the SAINT procedure and then corrected for absorption with SADABS procedure. The structures were solved by direct methods (SHELXT) and the refinement was done by full-matrix least squares on F2 (SHELX), using the Olex2 software package \cite{dolomanov_olex2_2009}.

\subsubsection{General Procedures for High-Throughput Synthesis of the AI-MOF series}\label{SI:procedure_synth}
High-throughput reactions were performed to screen synthesis conditions for all AI-MOFs using an automated process with Python scripts and a robotic liquid handler. First, the screening parameters, including concentration of the reagents, metal-to-linker ratio, amount of modulator, reaction temperature, and reaction time, were designed manually and converted into a CSV file that specified the information of the quantity of each reagent dispensed into each reaction vessel. Then, all reactions were carried out in 96-well plates, with each well containing a 0.5~mL borosilicate glass container. The programmed liquid handler, controlled by Python, dispensed mixtures of various stock solutions into the wells according to the CSV file. After dispensing, the plate was covered with a polytetrafluoroethylene sheet and sealed using a metal clamp before being heated in an isothermal oven at the designated reaction temperature and reaction time. Upon completion of the reaction, the plate was cooled and unsealed. The formation of MOFs in each well was examined under a microscope to identify suitable reaction conditions for the AI-MOFs.

\subsubsection{Database Curation and Training Set Formation}\label{databases}
Our training data combine experimental and computational MOF databases. For the experimental database, we used the CoRE MOF 2019 \cite{chung_advances_2019} dataset alongside a newly curated CSD database \cite{bruno_new_2002}, updated from version 5.39 (pre-November 2017) to version 5.45 (November 2023). Structures were first retrieved using ConQuest \cite{bruno_new_2002} from the 3D MOF subset (comprising non-disordered, three-dimensionally connected networks) and then cleaned with custom Python scripts to remove occupancy inaccuracies, missing hydrogen atoms, and both bound and unbound solvent molecules by the CSD Python API \cite{sykes_what_2024}. This processing yielded an additional 10,289 experimental MOF structures, which, when combined with computational databases (ultra MOF \cite{nandy_database_2023}, BW-DB \cite{boyd_data-driven_2019}, QMOF \cite{rosen_machine_2021}, ToBaCCo \cite{gomez-gualdron_evaluating_2016}) lead to a total of 500,000 MOFs. While training solely on experimental MOFs would improve the generation of synthesizable MOFs, the available dataset is too small for diffusion models, which require large databases—especially when handling complex MOF structures with a high atom count. In addition, computational databases is likely to lead to more exotic structures thus accelerating the exploration of more diverse MOF chemical space.

\subsubsection{Computational details of \emph{QForge} and \emph{QHarden}}\label{SI:DFT_wf}
The computational workflow was developed in the \texttt{atomate2} \cite{ganose2025atomate2} workflow orchestration package.
This package permits high-throughput calculation using a variety of computational methodologies, including interatomic potentials and electronic structure codes.
All MLFF calculations with MACE-MP-0 \cite{batatia_foundation_2024} and the GFN1-xTB \cite{grimme_robust_2017} tight-binding Hamiltonian were performed with \texttt{atomate2}'s interface to the atomic simulation environment (ASE) \cite{larsen2017ase}.

All electronic structure calculations were performed with plane wave DFT in the Vienna \textit{ab initio} simulation package (VASP) \cite{kresse_ab_1993,kresse1994vasp2,kresse_efficiency_1996,kresse_efficient_1996}.
We used two approximations for the exchange-correlation energy: the Perdew-Burke-Ernzerhof (PBE) generalized gradient approximation (GGA) \cite{perdew_generalized_1996} and r$^2$SCAN meta-GGA \cite{furness_accurate_2020}.
Although r$^2$SCAN captures some intermediate-range dispersion interactions implicitly, we augmented both PBE and r$^2$SCAN with an explicit D4 dispersion correction \cite{caldeweyher2019d4,ehlert2021r2scand4}.
An initial relaxation was performed with PBE-D4; converged wavefunctions were then used to initialize an r$^2$SCAN-D4 relaxation, and these converged wavefunctions were in turn used to compute an r$^2$SCAN-D4 single-point at the optimized geometry to obtain well-converged properties such as the electronic density of states.
We generally used the \texttt{MP24RelaxSet} as implemented in the \texttt{pymatgen} python package \cite{ong_python_2013}, but made simplifications to accommodate the complexity of these larger MOFs.
Total energies were converged within $10^{-5}$ eV; forces in the PBE-D4 relaxation within 0.1 eV/\AA{}, and forces within the r$^2$SCAN-D4 relaxation within 0.05 eV/\AA{}; we used a plane wave energy cutoff of 680 eV, and Gaussian smearing of the Fermi surface with width 0.05 eV.
Except in a few cases where a higher $k$-point density was used to initially test the sensitivity of the calculations to the $k$-mesh, only the $\Gamma$-point was used. The most recent ``PBE 64'' pseudopotentials (POTCARs) were used.

Because the diffusion-generated MOF structures were far from the DFT optimized structures, we pre-relaxed all structures using MACE-OMat-D4 (trained to the OMat24 database of PBE and PBE$+U$ calculations \cite{barrosoluque2024omat}).
All forces were converged within 0.1 eV/\AA{} in this step, as well as the MACE-MP-0 and GFN-xTB steps.

\subsubsection{Data Visualization and Accessibility}\label{SI:visualization}
The database, including detailed computed properties, is made publicly available through the Materials Project website (MP contrib section). Each component of the workflow can be found in \texttt{atomate2}.


\subsection{Supplementary Text}

\subsubsection{\emph{LinkerGen}'s in-context learning strategy for chemical composition generation} \label{SI:Linker_Gen}
To generate chemically valid and stoichiometrically accurate chemical formulae, we used LLM strategies. In contrast to random generation, LLMs have demonstrated the capability to learn and generate complex molecular representations, through the SMILES representation, allowing exploration of novel molecules \cite{flam-shepherd_language_2022, m_bran_augmenting_2024, cavanagh_smileyllama_2024}. Our aim was to produce chemical formulae that resemble experimentally known organic linkers while still introducing sufficient novelty to explore the unreported regions of the MOF chemical space. SMILES representation was selected for its ease of rapid analysis and wide availability in software and Python libraries. Experimental organic linker representations were provided as chemical formulae or SMILES. Furthermore, we used chain-of-thought prompting strategies \cite{wei_chain--thought_2023} to impose constraints on chemical formula generation, ensuring correct stoichiometric ratios as well as metal SBUs. To further enhance structural diversity, supercell formulations were generated by scaling the total number of atoms, organic linkers and metal SBUs, with a maximum of 256 atoms.

\begin{figure}
    \centering
    \includegraphics[width=1.0\textwidth]{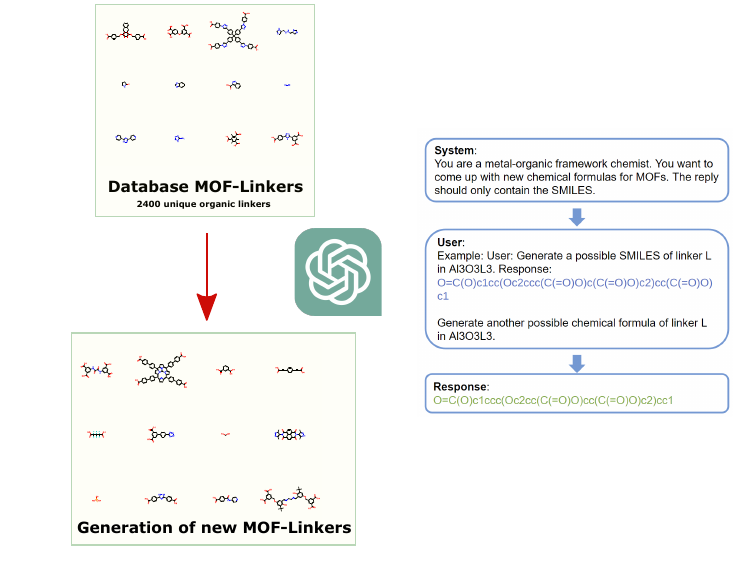}
    \caption{\textbf{Example of prompt used for the organic linker SMILES generation.} A general prompt giving context of the MOF field is given to the system prompt. The user prompt include the experimental MOF organic linkers and the kekulized validated SMILES are then stored in a CSV file and convert to chemical formulae.}
    \label{fig:SI_LinkerGen_prompt}
\end{figure}

\begin{figure}
    \centering
    \includegraphics[width=1.0\textwidth]{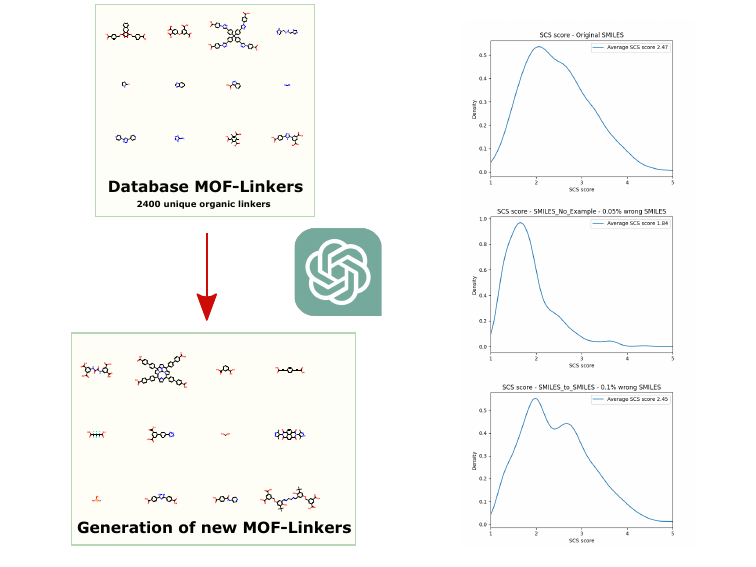}
    \caption{\textbf{Distribution of SCScore for generated organic linkers with different strategies.} At the top, the distribution of the experimental MOF organic linkers. Middle, the SMILES generated without in-context learning. Bottom, SMILES generated throught in-context learning and chain-of-thought. }
    \label{fig:SI_LinkerGen_score}
\end{figure}

An example of prompt for our in-context learning strategy is provided in Fig.~\ref{fig:SI_LinkerGen_prompt}. To evaluate chemical diversity and reliability relative to experimental linkers, we used synthesizability scores and chemical composition analysis. As first attempt, we used a zero-shot generation strategy without linker context that resulted in chemical formulae heavily biased towards commonly reported linkers, such as the widely used MOF-5 organic linker \cite{rosi_hydrogen_2003}, as shown in Fig.~\ref{fig:SI_LinkerGen_score}. The SCScore analysis under zero-shot conditions exhibited a narrow peak at 1.84 (Fig.~\ref{fig:SI_LinkerGen_score} b)), lower than the broader experimental distribution centered around 2.47 (Fig.~\ref{fig:SI_LinkerGen_score} a)). In comparison, when our in-context learning approach combined with chain-of-thought prompts was used, the resulting SCScore distribution was similar to the one from the experimental data, with an average value of 2.47 and wide peaks with scores of 2 and 3, which suggests that this strategy help to generate linkers that are chemically valid.

However, this results also shows the limitation of using only LLM for the discovery of MOF chemistry, since LLM will generate distribution of linkers very similar to the ones given in the prompt and will slightly mixed them together by adding some functional groups or through atom substitution. A similar trend was found in Fig.~\ref{fig:figure3} a), although the organic linkers generated by LLM alone exhibit a close distribution overlap compared to those obtained through the combined LLM and diffusion model, the latter occupy a distinct region of the chemical space. While we envisioned that it is possible to push the exploration of the chemical space through careful prompting with LLM, these findings highlight the importance and potential of combining LLMs with diffusion-based models to achieve a more broader and more diverse exploration of chemical space.

\newpage

\subsubsection{Computational details of the Diffusion model and MOF representation of \emph{CrystalGen}.} \label{SI:diffusion}
Typical MOF systems are much larger than inorganic crystals, MOF systems contain hundreds to thousands of atoms, whereas typical inorganic crystals often contain less than a dozen atoms in a unit cell. As a result, generative models designed for inorganic crystals such as UniMat~\cite{yang_scalable_2024} and MatterGen~\cite{zeni_generative_2025} cannot be directly applied to modeling MOF systems.

To address these limitations, we represent a MOF structure as a point cloud in a similar spirit to Yang et al.\cite{yang_generative_2024}. Specifically, a MOF system with $A$ atoms is represented by a matrix of shape $[A, 4]$, where the first three dimensions of each atom represent the $x,y,z$ coordinates of each atom and the last dimension denotes the atomic number. We use $A=256$ to cap the maximum number of atoms we model. The atomic number is represented as a normalized continuous value within the diffusion model’s input range. In the diffusion model, each atom is processed through a multi-layer perceptron (MLP), followed by order-invariant self-attention across atoms (without positional encoding). Unlike the typical U-Net architecture used for image generation, this approach avoids resolution-changing downsampling or upsampling passes. However, we retain the concatenated skip connections commonly employed in U-Net architectures to preserve information flow. While preserving the resolution (number of points), we repurpose the spatial downsampling and upsampling passes typically used in U-Net architectures for images or videos. Additionally, we employ residual networks with concatenated skip connections.

\newpage

\subsubsection{Computational details of \emph{QForge}}\label{SI:qforge}

Candidate structures processed by \emph{QForge} are first screened with Zeo++, under the assumption that a valid MOF must (i) exhibit a pore-limiting diameter greater than 2.5,\AA{}, size of the $N_{2}$ sorbate, (ii) possess a probe-occupiable accessible volume fraction above 30$\%$, and (iii) have an accessible volume exceeding the non-accessible volume. The filtered structures are then geometry-optimized using MACE-MP-0, with an $f_\mathrm{max}$ threshold of 0.1, using the  library \cite{ganose2025atomate2}. An additional screening step with Zeo++, applying the same criteria is performed. Finally, a last geometry optimization is carried out using the semi-empirical GFN1-xTB method.

To determine the most suitable semi-empirical model for relaxing MOF structures, we evaluated the convergence success rate, cell parameters, and computational time. All calculations were performed using \texttt{TBLite} with an $f_\mathrm{max}$ threshold of 0.1. We randomly selected 125 MOF samples from the QMOF database and assessed the performance of both GFN1-xTB \cite{grimme_robust_2017} and GFN2-xTB \cite{bannwarth_gfn2-xtbaccurate_2019}. 

GFN1-xTB demonstrated a higher convergence success rate compared to GFN2-xTB, while both functionals showed similar behavior in terms of structural relaxation metrics, such as cell parameters (see Table \ref{tabSI:GFN-xTB_analysis}). We chose to use GFN1-xTB due to its higher convergence success rate, critical for structure that might be far from equilibrium like in diffusion. 

A prior study applied GFN1-xTB on the entire CoRE MOF database \cite{chung_advances_2019} effectively preserving the crystal structures of MOFs, with good agreement to experimental cell parameters and bonding geometries \cite{nurhuda_performance_2022}. In addition, a recent benchmarking effort \cite{thi_le_evaluation_nodate} evaluated various semi-empirical quantum method, including GFN1-xTB on UiO-66 \cite{cavka_new_2008} against highly accurate hybrid DFT. Across a range of properties (e.g, ligand exchange reactions, host-guest interactions), GFN1-xTB consistently outperformed other models, including GFN2-xTB.

Our results support these recent finding and further confirm that GFN1-xTB is a robust and efficient choice for large-scale, high-throughput screening computations of MOF and can be further enhanced when used in tandem with MLFF such as MACE-MP-0 here.

\newpage

\subsubsection{Computational details of \emph{SynthABLE}}\label{SI:synthable}

We performed a compositional analysis of the organic linkers extracted from diffusion-generated MOFs (Fig.~\ref{figSI:composition_linker_synthable}). Our first observation is that the large majority of these linkers are novel, with only 73 previously experimentally reported. A large fraction of the linkers possess two carboxylate groups, a common feature in zinc-based MOFs due to the strong coordination preference of zinc for bidentate carboxylates. Additionally, most linkers are aromatic and do not contain pyridine, consistent with experimental trends.

These findings provide insights that could inform future generative models for synthesizable linkers. In particular, the observed prevalence of certain chemical features (e.g., carboxylate-rich, aromatic) may guide the design of synthetic heuristics or inspire the exploration of underrepresented linker types, such as highly carboxylated or pyridine-enriched linkers, which remain relatively rare but may be synthetically viable.

To complement our analysis, we conducted a blind survey, in which five experimentalists were presented with 20 unique organic linkers. Each set included 8 linkers predicted to be unsynthesizable by the Allchemy software. This human-in-the-loop evaluation enabled us to assess agreement between expert intuition and ML-based predictions software in an unbiased setting.

To further narrow the candidate pool for experimental synthesis, we applied a set of filtering criteria for synthesizability accessibility: (i) single-linker only MOFs (i.e., no mixed-linker), (ii) linkers not previously reported in experimental databases, (iii) linkers containing at least two carboxylate groups, (iv) linkers with partially or fully aromatic backbones (i.e., excluding fully aliphatic species), and (v) MOFs incorporating only known and already synthesized metal SBUs (e.g., paddlewheel, tetrahedral motifs).

\newpage

\subsubsection{Predictive capabilities on different metal SBUs}\label{SI:formation_energies}

To evaluate the predictive capability and generalizability of our system of agents, we generated 20,000 MOFs with different metal SBUs composition based on Aluminium (Al), Magnesium (Mg), and Zinc (Zn), distinct from the database analyzed in the main section. These metals were selected due to their widespread synthesis in MOFs and relative ease of synthesis. The compositions were tailored based on known MOFs, such as MOF-303 and MOF-801 (known for water harvesting), as well as MOF-74 and IRMOF-74 (known for carbon capture), while ensuring that the total number of atoms per structure remained below 256, as explained in Section \ref{SI:diffusion}. Around 2,000 MOFs successfully passed all our screening criteria, from which we selected 60 structures for extensive DFT optimization using the procedure mentioned previously Formation enthalpy analyses revealed that nearly all optimized structures exhibited negative enthalpy values, indicating thermodynamic stability, Fig.~\ref{figSI:formation_Mg_Al_Zn}. Notably, structures containing zinc SBUs demonstrated higher stability, consistent with their greater use in MOFs, and thus more present in the training set of the diffusion model, followed by magnesium and aluminium SBUs. These results demonstrate our model applicability across diverse MOF chemistry, marking a significant step towards the development of foundation models for MOF discovery.

\newpage

\subsubsection{Analysis of MOF crystal structures selected by \emph{QHarden}.}\label{SI:mace-mp0_analysis}

Figures \ref{fig:figure3}a and \ref{fig:figure3}b show the distribution of thermal decomposition temperature (T$_{d}$) and solvent removal stability on the QMOF dataset~\cite{rosen_machine_2021}, a subset of QMOF and our diffusion-generated MOFs. To enable meaningful comparison, the subset of the QMOF dataset was constructed in order to matches the elemental composition, topology and known zinc SBUs (e.g. paddlewheel, tetrahedral motifs) of our diffusion-generated MOFs. The predictions were obtained using the machine learning predictor developed by Nandy \textit{et al.}\cite{nandy_mofsimplify_2022}. The mean and standard deviation of each distribution can be found in Table \ref{tabSI:bulk_modulus}

The QMOF subset show a slight peak shift toward higher solvent removal stability that can be explain by the higher stability of zinc-based MOF compared to other metal and MOF-5 topology, the predicted solvent removal stability of MOF-5 being 0.92. Similarly, the decomposition temperature is shifted toward higher decomposition temperature that also goes toward MOF-5 predicted decomposition temperature of around 410$^\circ$C that collaborate with our finding that the model tend to generate MOF-5 topologies. We also find that the predicted decomposition temperature of MOF-5 is within the error bar of the experimental values that are between 450-500$^\circ$C \cite{healy_thermal_2020}. Compared with the ``ultra''-stable database of Nandy  \textit{et al.}\cite{nandy_database_2023} our diffusion-generated MOFs exhibit a higher stability with a very high narrow peak located at 0.88 for the solvent removal stability while having a decomposition temperature close to the QMOF subset. 

While several studies have used MLFF to evaluate mechanical properties of inorganic crystals, their application to MOFs has been very limited \cite{elena_machine_2024, eckhoff_molecular_2019, wieser_machine_2024}. Due to the lack of reference datasets, both experimental and computed ones such as DFT. Early work by Eckhoff \textit{et al.} first show that MLFFs could predict DFT-level bulk modulus of MOF-5.

Recently, a fine-tuned MLFF, MACE-MP-MOF0, based on the MACE-MP-0b foundation model \cite{batatia_foundation_2024}, was shown to achieve prediction in-par with experimental bulk moduli for a set of four MOFs. In several cases, MACE-MP-MOF0 predictions aligned more closely with experimental values compared to DFT, due to possible failure in the geometry optimization and lack of statistic. These findings highlight the potential of MLFFs as both accurate and computationally efficient tools for high-throughput screening.

In this study, we used MACE-MP-0b to calculate the bulk modulus on the whole QMOF database. For each MOF, a linear strain of $\pm 5\%$ was applied to the relaxed geometry, and ionic positions were optimized at fixed cell volume and then fit to the Vinet equation of state (EOS) to extract the zero-temperature bulk modulus. This high-throughput workflow used our latest \texttt{atomate2} framework. For the QMOF database, 99.69\% (20,311/20,375) bulk modulus were successfully calculated, within the 0--80~GPa range. While for our Diffusion+LLM database, 95.65\% (3,675/3,842) yielded valid bulk modulus values. A small number of EOS fittings failed to converge, likely due to MACE-MP-0b limitations in extreme volumetric strain.

These results demonstrate, for the first time that MLFFs can be used at scale to screen mechanical properties of MOFs, enabling deeper understanding of structural stability, and mechanical performance of MOF materials.

\newpage

\subsubsection{Synthesis of the zinc muconate MOF, AI-MOF-1}\label{SI:synth_AI-MOF-1}
The trans,trans-muconic acid (3.8 mg, 0.027 mmol) and zinc nitrate hexahydrate (23.8 mg, 0.080 mmol) in 10 mL of $N,N$-diethylformamide were dissolved in a 20-mL vial. Then, cap the vial tightly and heat it at 100 $^oC$ for 24 h. Then washed it with $N,N$-dimethylformamide for 3 times.

\subsubsection{Synthesis of the perylene-based MOF, AI-MOF-2}\label{SI:synth_AI-MOF-2}
In a 20~mL scintillation vial, dissolve 10~mg of perylene-3,9-dicarboxylic acid and 30~mg Zn(NO$_{3}$)$_{2}$ $\cdot$ 6H$_{2}$O in 5~mL \textit{N, N}-diethylformamide (DEF) using sonication. The solution has been heated up in a 120$^{\circ}$C oven for 16~h to grow red block-shape crystals. The crystals have been washed with DMF × 3 (10~mL) and acetone × 3 (10~mL) and the PXRD has been collected on the wet sample. 
The AI-MOF-2 exhibit the same topology as iconic MOF-5 due to the linear, 180$^{\circ}$ angle, between carboxylate groups. 

\subsubsection{Synthesis of the zinc dimethyl fumarate MOF with DEF solvent, AI-MOF-3}\label{SI:synth_AI-MOF-3}
In a 35 mL microwave vial sealed with a silicone cap, dissolve 59 mg (0.200 mmol) of Zn(NO$_{3}$)$_{2}$ $\cdot$ 6H$_{2}$O and 14 mg (0.100 mmol) of 2,3-Dimethylfumaric acid in 10 mL of Diethylformamide. The solution was heated to 100$^{\circ}$C in a microwave reactor for 1 h to grow light-green colored cube-shaped crystals, Fig.\ref{fig:synth_AI-MOF-3_crys}. The resulting mother liquor was measured using PXRD in its wet form, Fig.\ref{fig:synth_AI-MOF-3_pxrd}.

\subsubsection{Synthesis of the zinc dimethyl fumarate MOF with DMF solvent, AI-MOF-4}\label{SI:synth_AI-MOF-4}
In a 20~mL scintillation vial, dissolve 50~mg of 2,3-dimethylfumaric acid and 150~mg Zn(NO$_{3}$)$_{2}$ $\cdot$ 6H$_{2}$O in 10~mL DMF. The solution was heated in a 120$^{\circ}$C oven for 16~h to grow colorless block-shaped crystals, Fig.\ref{fig:synth_AI-MOF-4_crys}. The crystals were washed with DMF and the PXRD was collected on the wet sample, Fig.\ref{fig:synth_AI-MOF-4_pxrd}. The crystallographic data can be found in Table \ref{tabSI:SM621B}. checkCIF/PLATON report for the crystallographic data, namely AI-MOF-4, with no syntax errors were detected.

\subsubsection{Synthesis of the aluminum dimethyl fumarate MOF, AI-MOF-5} \label{SI:synth_AI-MOF-5}
Equal molar of 2,3-Dimethyl-2-butenedioic acid and aluminum chloride hexahydrate were dissolved in a DMF–water mixture (1/3, v/v). The solution was sealed and heated in a 100 $^{\circ}$C oven for 24 h to obtain products as white solids, Fig.\ref{fig:synth_AI-MOF-5_crys}. PXRD was collected on the wet sample and the product showed a primary peak at 2$\theta$ = 7.5$^\circ$, consistent with a crystalline MOF phase Fig.\ref{fig:synth_AI-MOF-5_pxrd}.


\begin{figure}[ht]
    \centering
    \includegraphics[width=\textwidth]{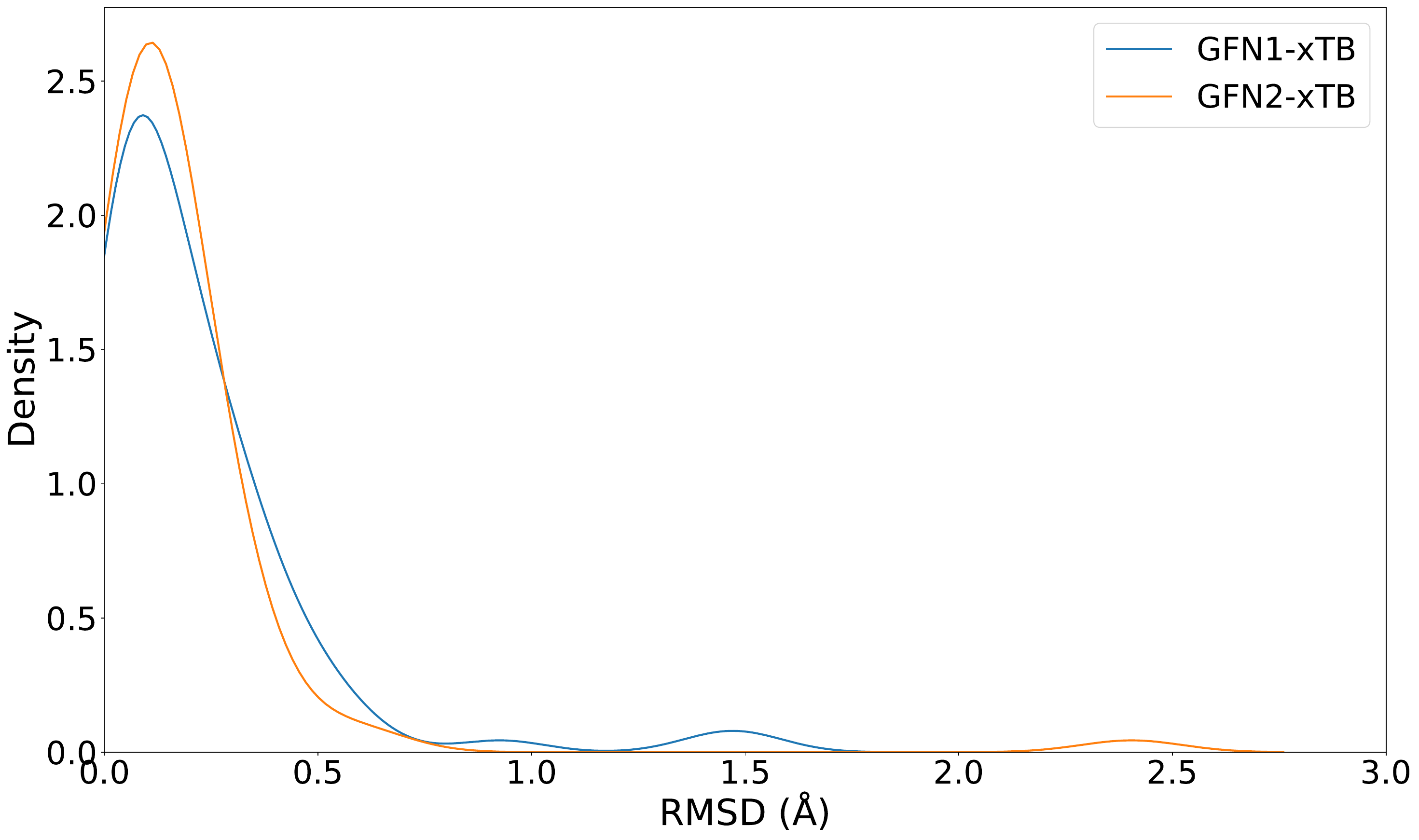}
    \caption{\textbf{Root mean square deviation (RMSD) comparison between GFN1-xTB and GFN2-xTB.} The RMSD was computed between the initial DFT optimized structure from QMOF and the optimized one with each method.}
    \label{figSI:GFN-xTB_analysis}
\end{figure}

\begin{figure}[ht]
    \centering
    \includegraphics[width=0.7\textwidth]{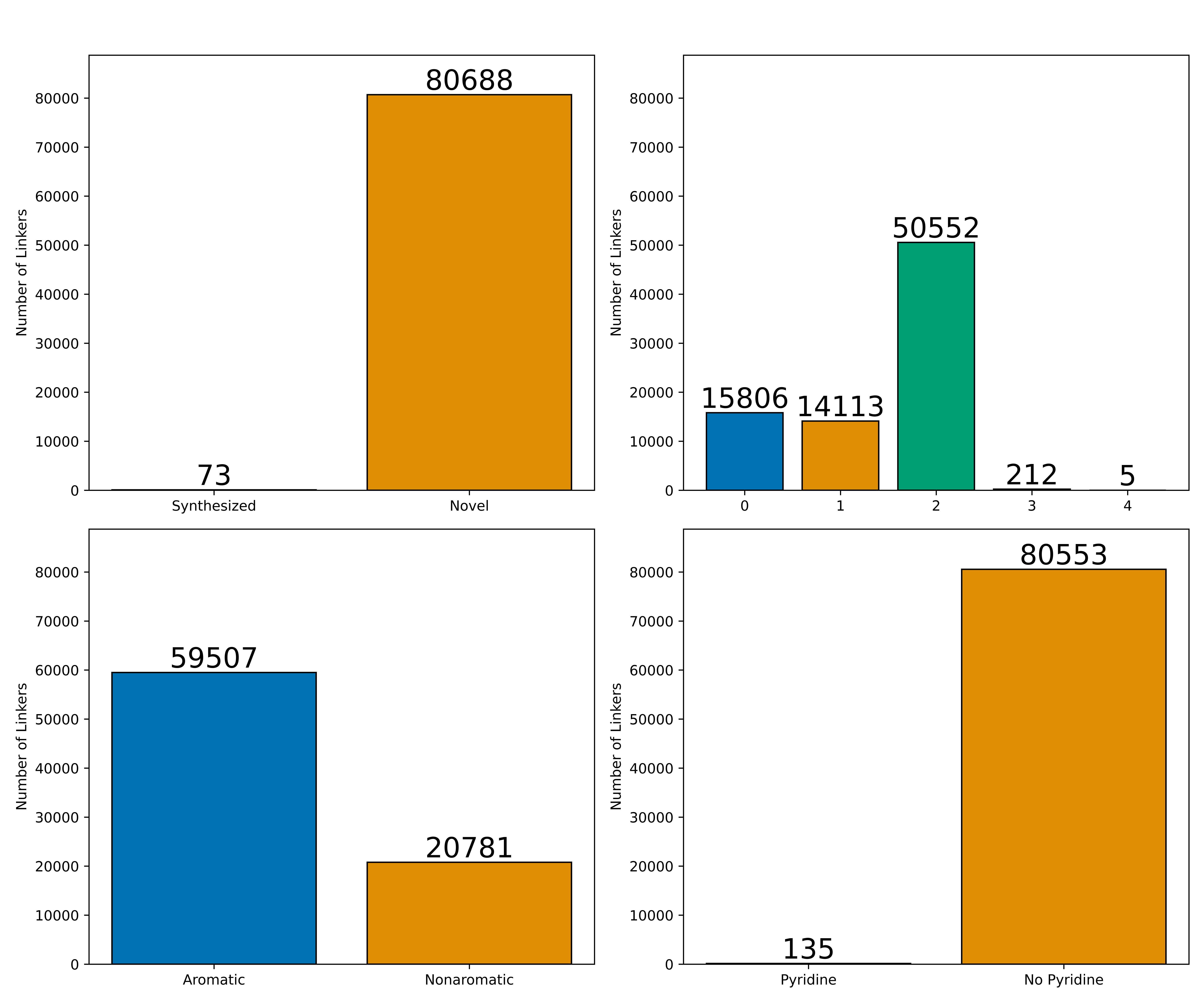}
    \caption{\textbf{Chemical composition analysis of MOF linkers generated with known Zn secondary building units (SBUs).}
    (\textbf{Top panels:}) (left) total number of novel linkers; (right) distribution of linkers by the number of carboxyl groups.
    (\textbf{Bottom panels:}) (left) number of aromatic linkers; (right) distribution of linkers containing pyridine.}
    \label{figSI:composition_linker_synthable}
\end{figure}

\begin{figure}[ht]
    \centering
    \includegraphics[width=0.7\textwidth]{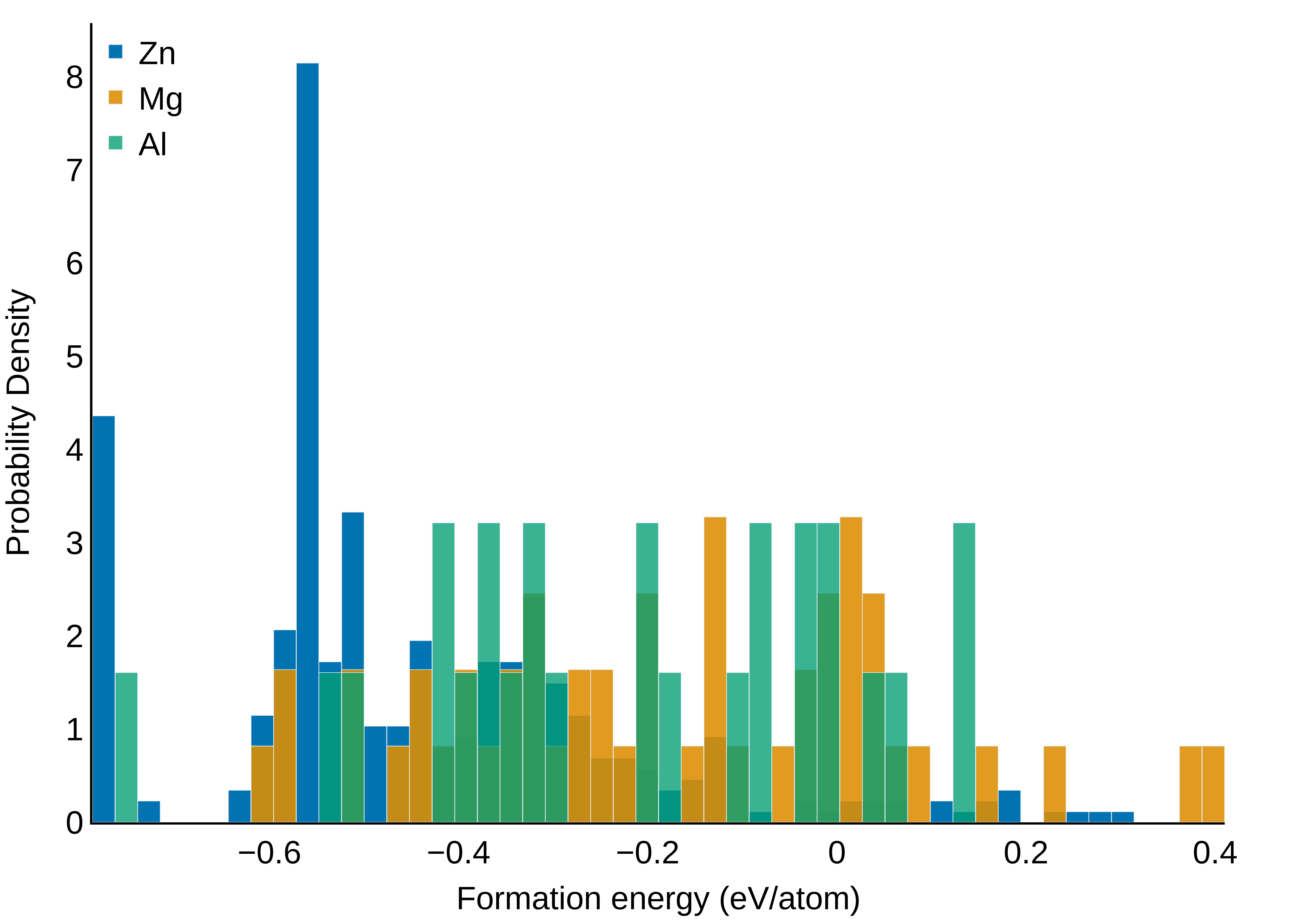}
    \caption{\textbf{Formation energy distribution of diffusion-generated MOFs with various metal SBUs.} The formation energies of generated MOFs containing aluminium, magnesium, and zinc SBUs were computed using the r$^2$SCAN-D4 functional, following the procedure described in Section \ref{SI:DFT_wf}.}
    \label{figSI:formation_Mg_Al_Zn}
\end{figure}

\begin{figure}[ht]
  \centering
  \includegraphics[width=0.7\textwidth]{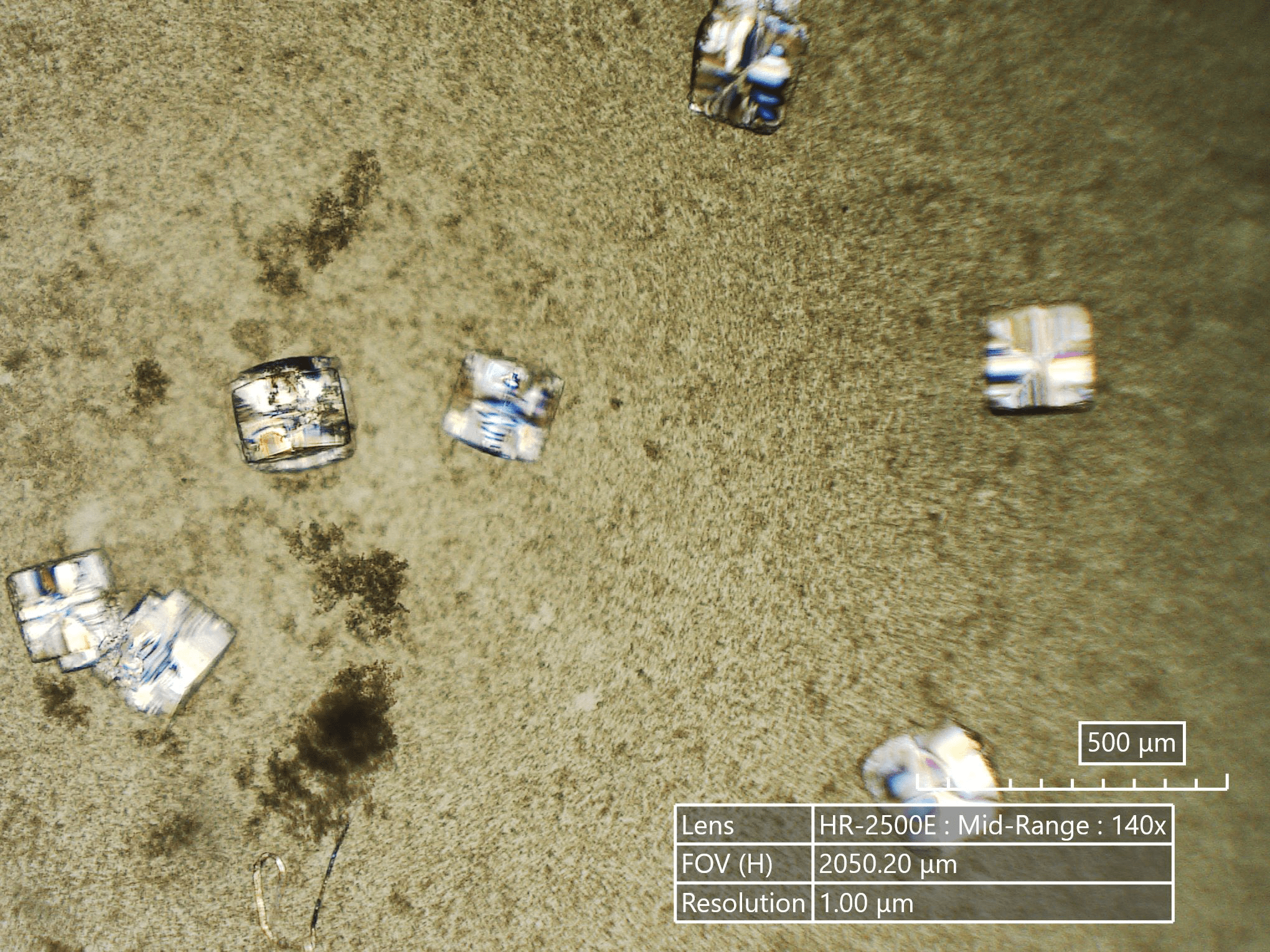}
  \caption{\textbf{Representative optical image of the zinc muconate MOF, AI-MOF-1.}}
  \label{fig:synth_AI-MOF-1_crys}
\end{figure}

\begin{figure}[ht]
  \centering
  \includegraphics[width=0.8\textwidth]{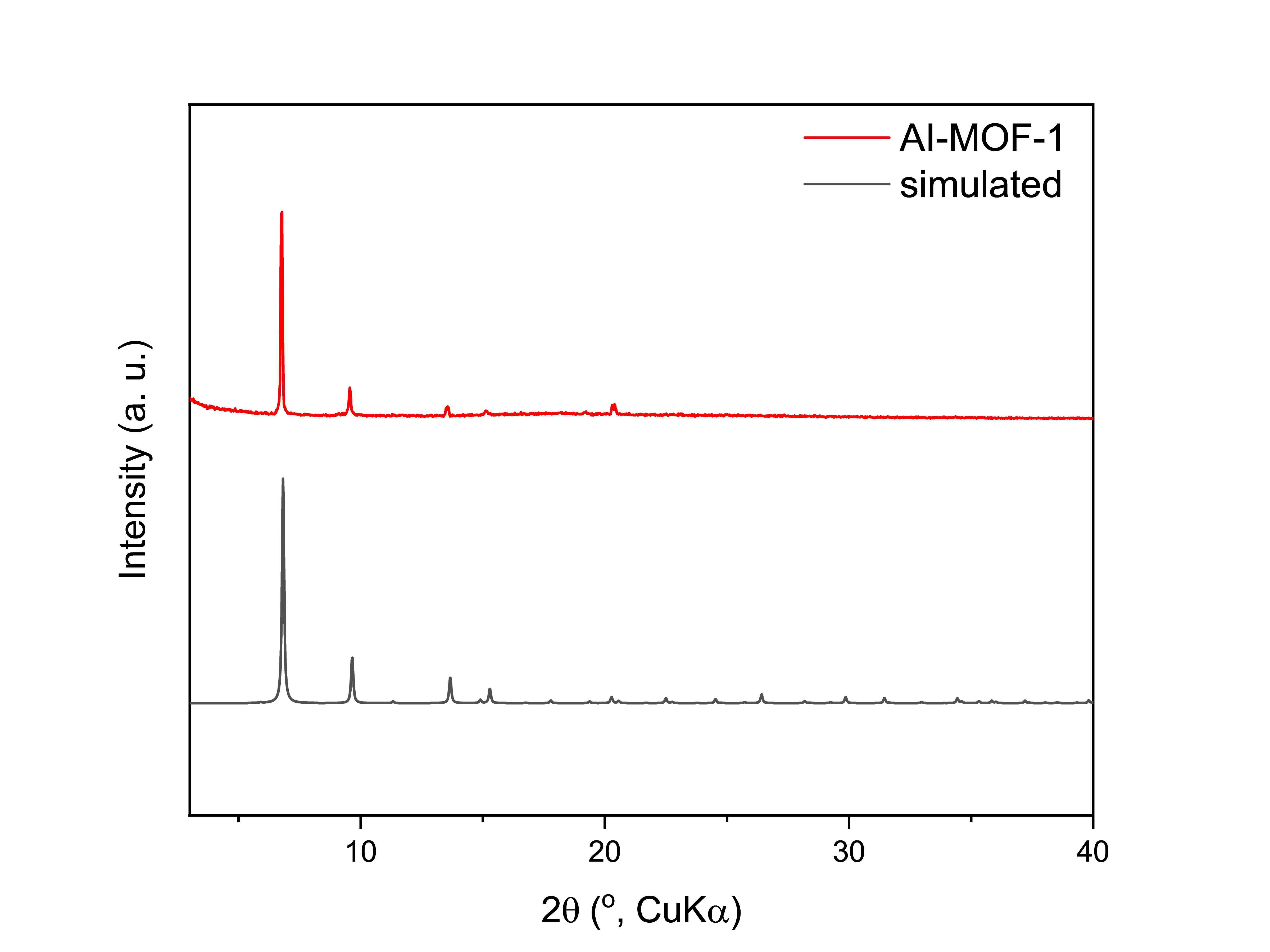}
  \caption{\textbf{Experimental and simulated PXRD of the zinc muconicate MOF, AI-MOF-1.}}
  \label{fig:synth_AI-MOF-1_pxrd}
\end{figure}

\begin{figure}[ht]
  \centering
  \includegraphics[width=0.7\textwidth]{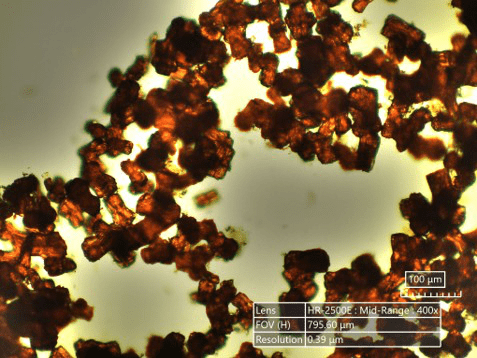}
  \caption{\textbf{Representative optical image of the perylene-based MOF, AI-MOF-2.}}
  \label{fig:synth_AI-MOF-2_crys}
\end{figure}

\begin{figure}[ht]
  \centering
  \includegraphics[width=0.8\textwidth]{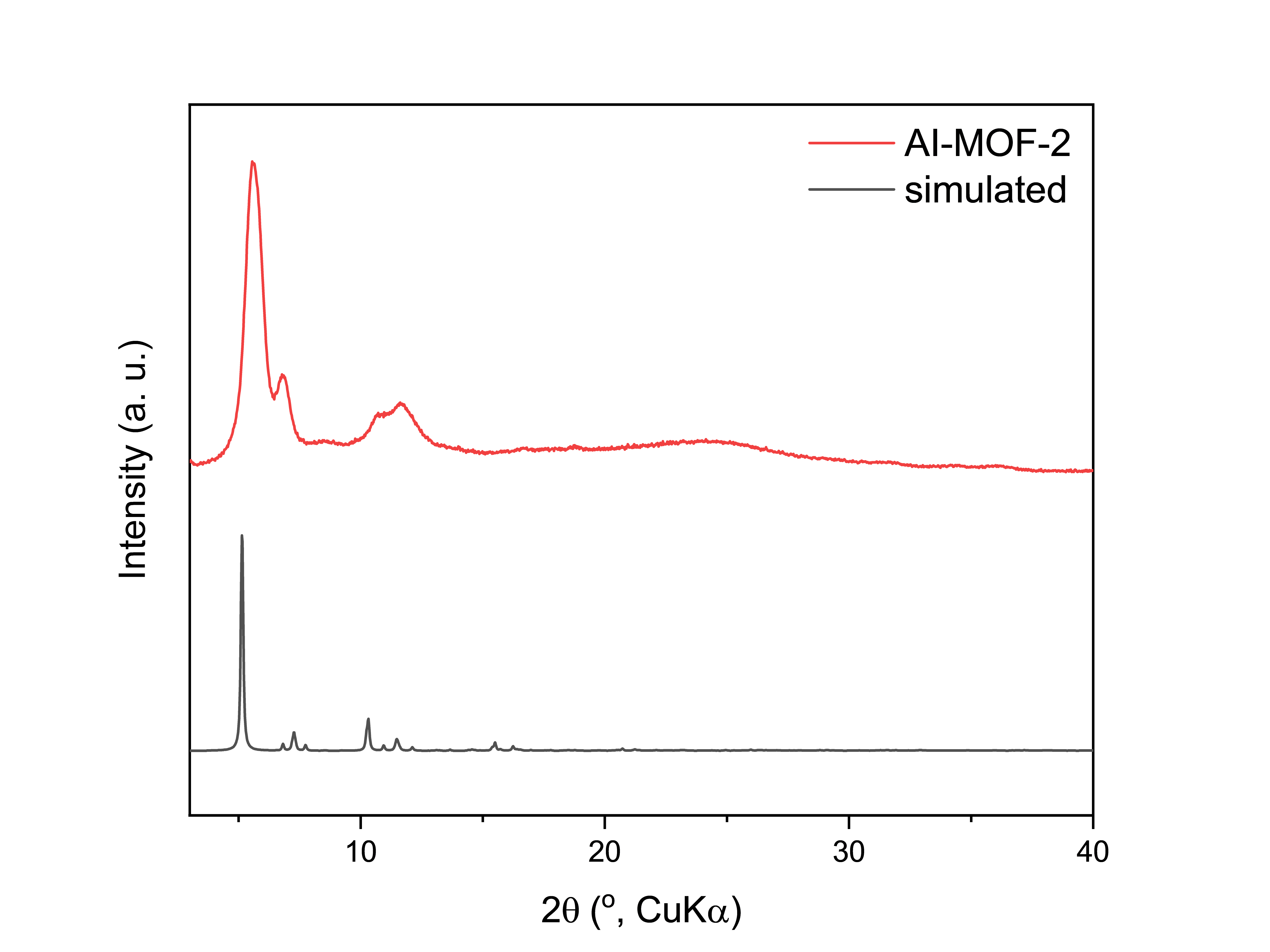}
  \caption{\textbf{Experimental and simulated PXRD of the perylene-based MOF, AI-MOF-2.}}
  \label{fig:synth_AI-MOF-2_pxrd}
\end{figure}

\begin{figure}[ht]
  \centering
  \includegraphics[width=0.7\textwidth]{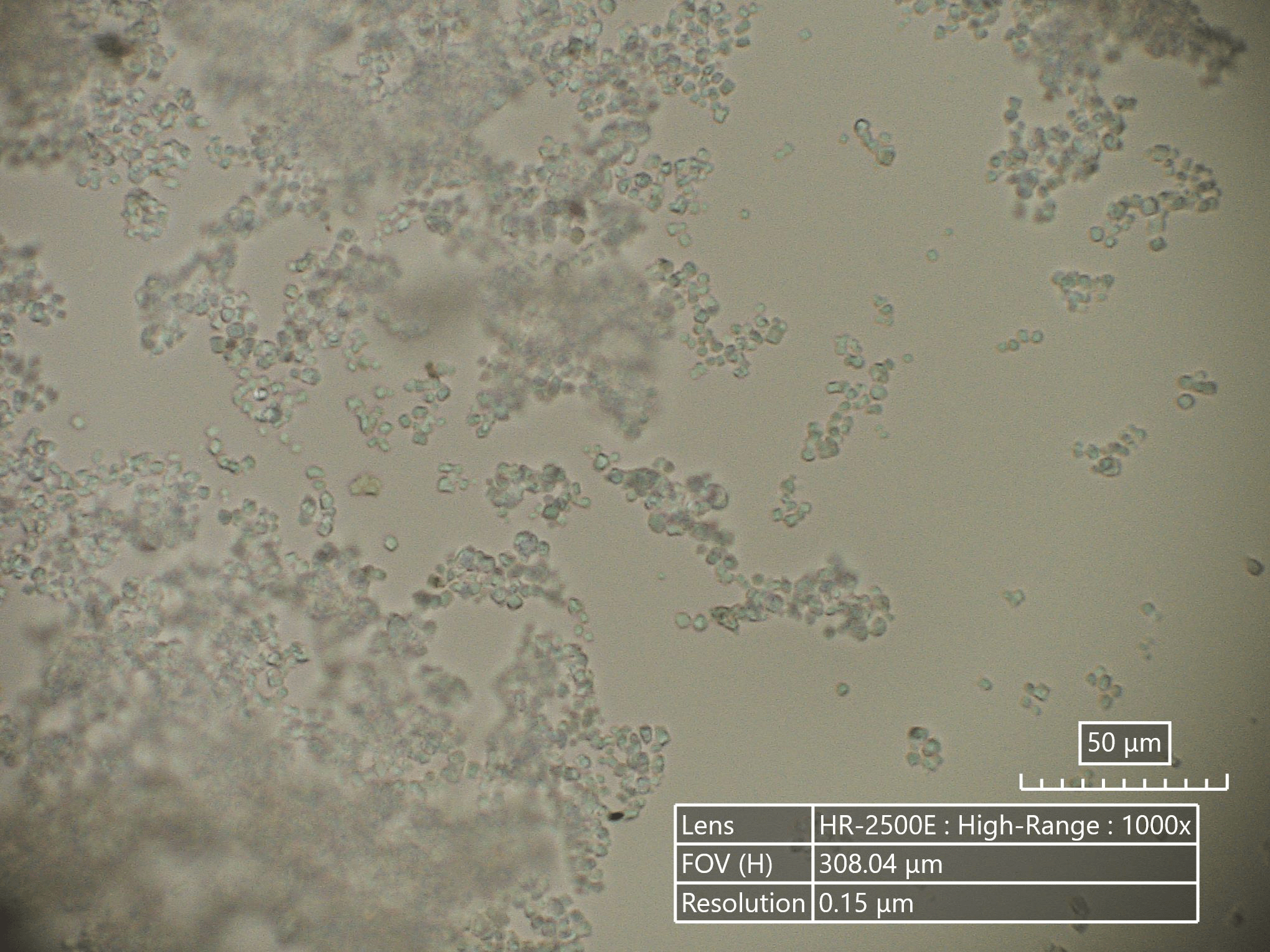}
  \caption{\textbf{Representative optical image of the zinc dimethyl fumarate MOF with DEF solvent, AI-MOF-3}}
  \label{fig:synth_AI-MOF-3_crys}
\end{figure}

\begin{figure}[ht]
  \centering
  \includegraphics[width=0.8\textwidth]{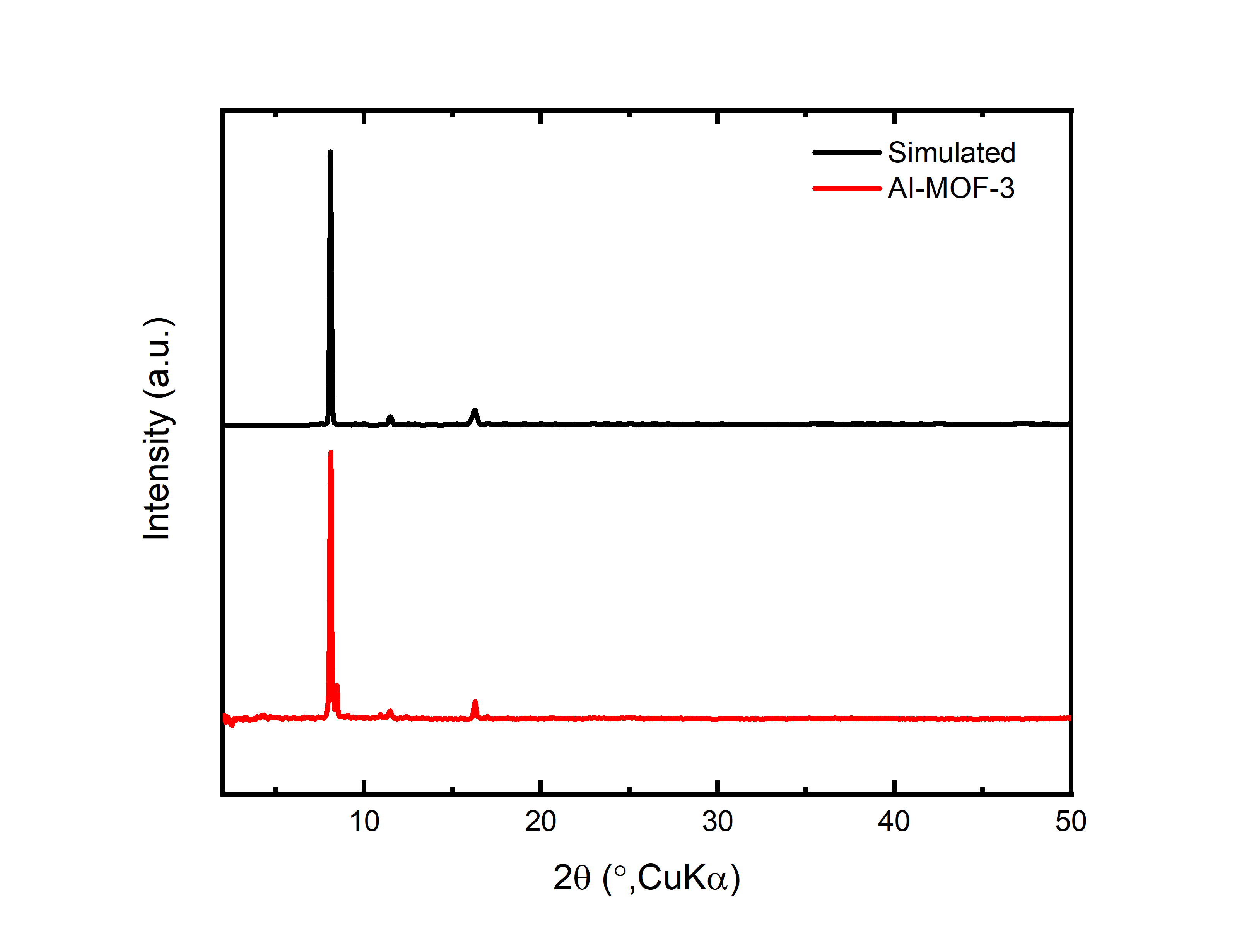}
  \caption{\textbf{Experimental and simulated PXRD of the zinc dimethyl fumarate MOF with DEF solvent, AI-MOF-3}}
  \label{fig:synth_AI-MOF-3_pxrd}
\end{figure}

\begin{figure}[ht]
  \centering
  \includegraphics[width=0.7\textwidth]{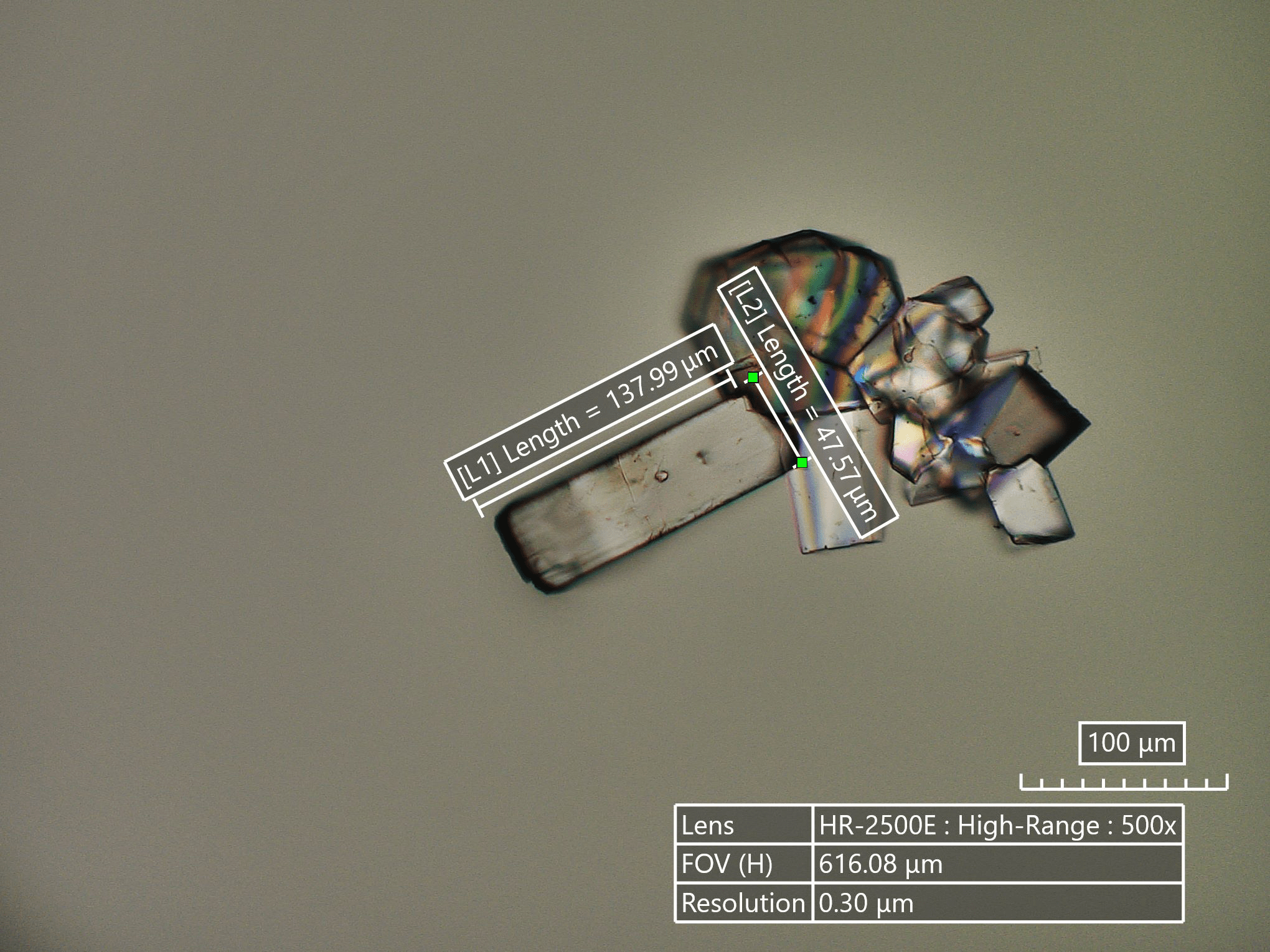}
  \caption{\textbf{Representative optical image of the zinc dimethyl fumarate MOF with DMF solvent, AI-MOF-4.}}
  \label{fig:synth_AI-MOF-4_crys}
\end{figure}

\begin{figure}[ht]
  \centering
  \includegraphics[width=0.8\textwidth]{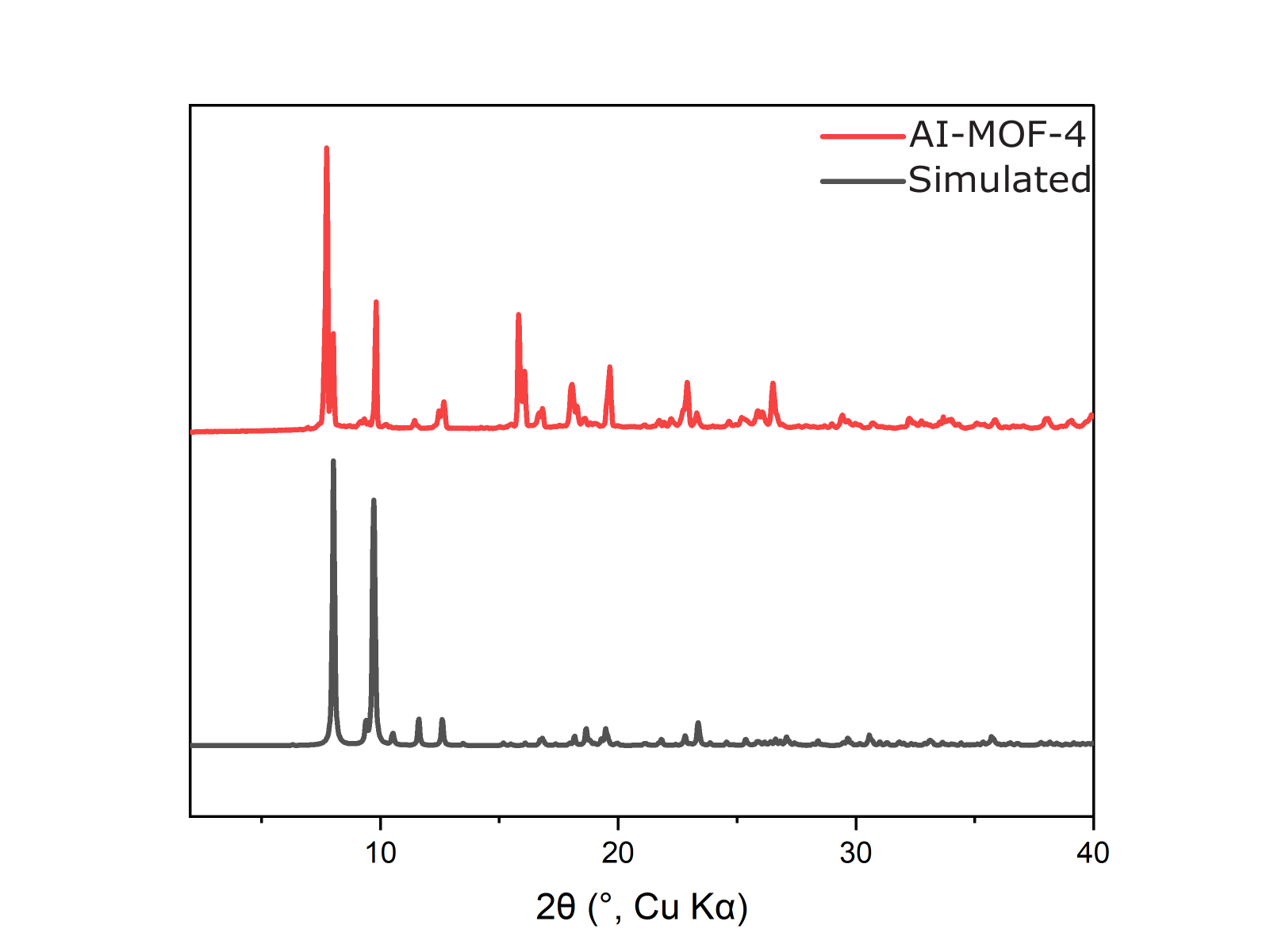}
  \caption{\textbf{Experimental and simulated PXRD of the zinc dimethyl fumarate MOF with DMF solvent, AI-MOF-4.}}
  \label{fig:synth_AI-MOF-4_pxrd}
\end{figure}

\begin{figure}[ht]
  \centering
  \includegraphics[width=0.7\textwidth]{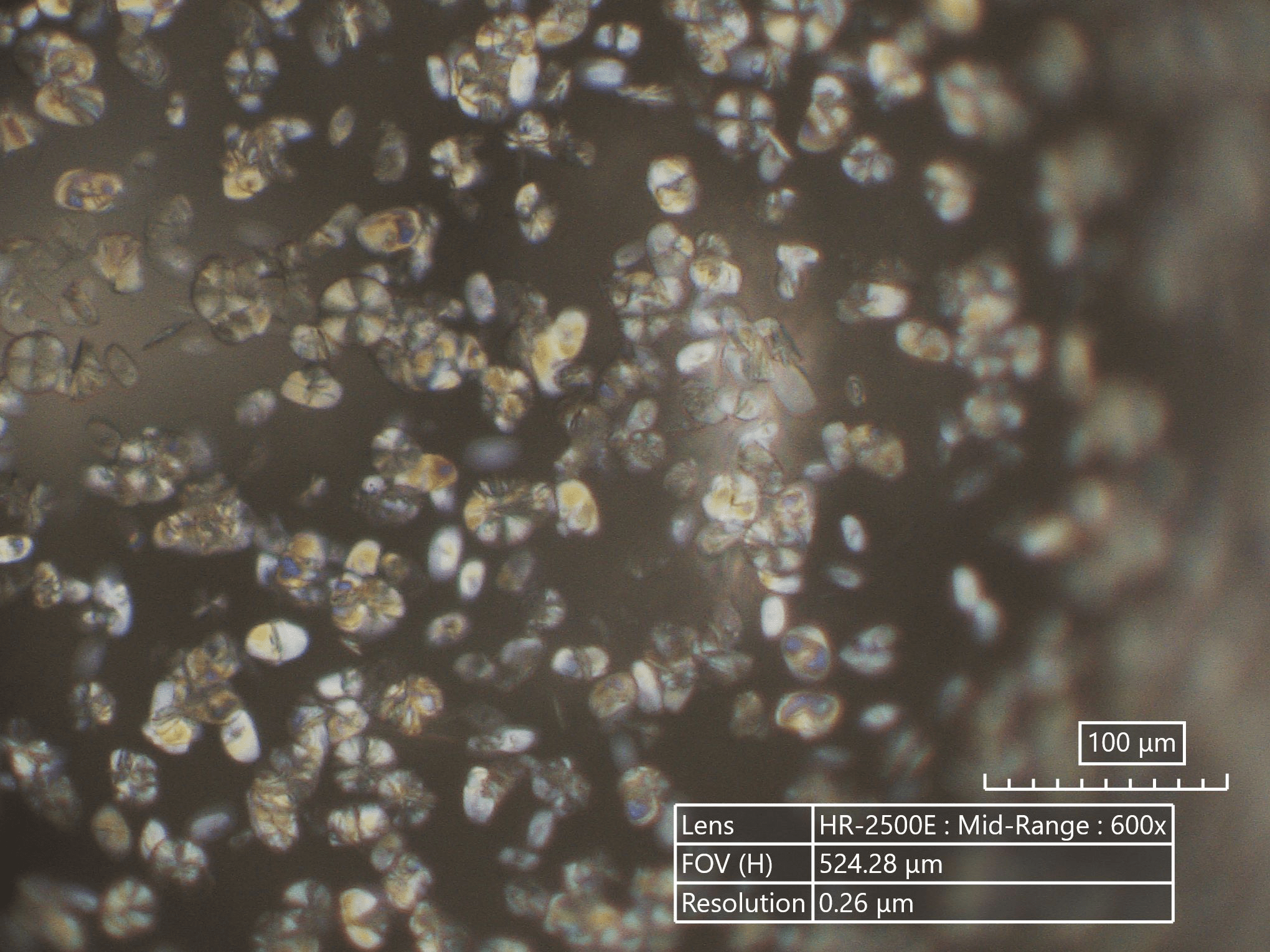}
  \caption{\textbf{Representative optical image of the aluminum dimethyl fumarate MOF with DMF solvent, AI-MOF-5.}}
  \label{fig:synth_AI-MOF-5_crys}
\end{figure}

\begin{figure}[ht]
  \centering
  \includegraphics[width=0.8\textwidth]{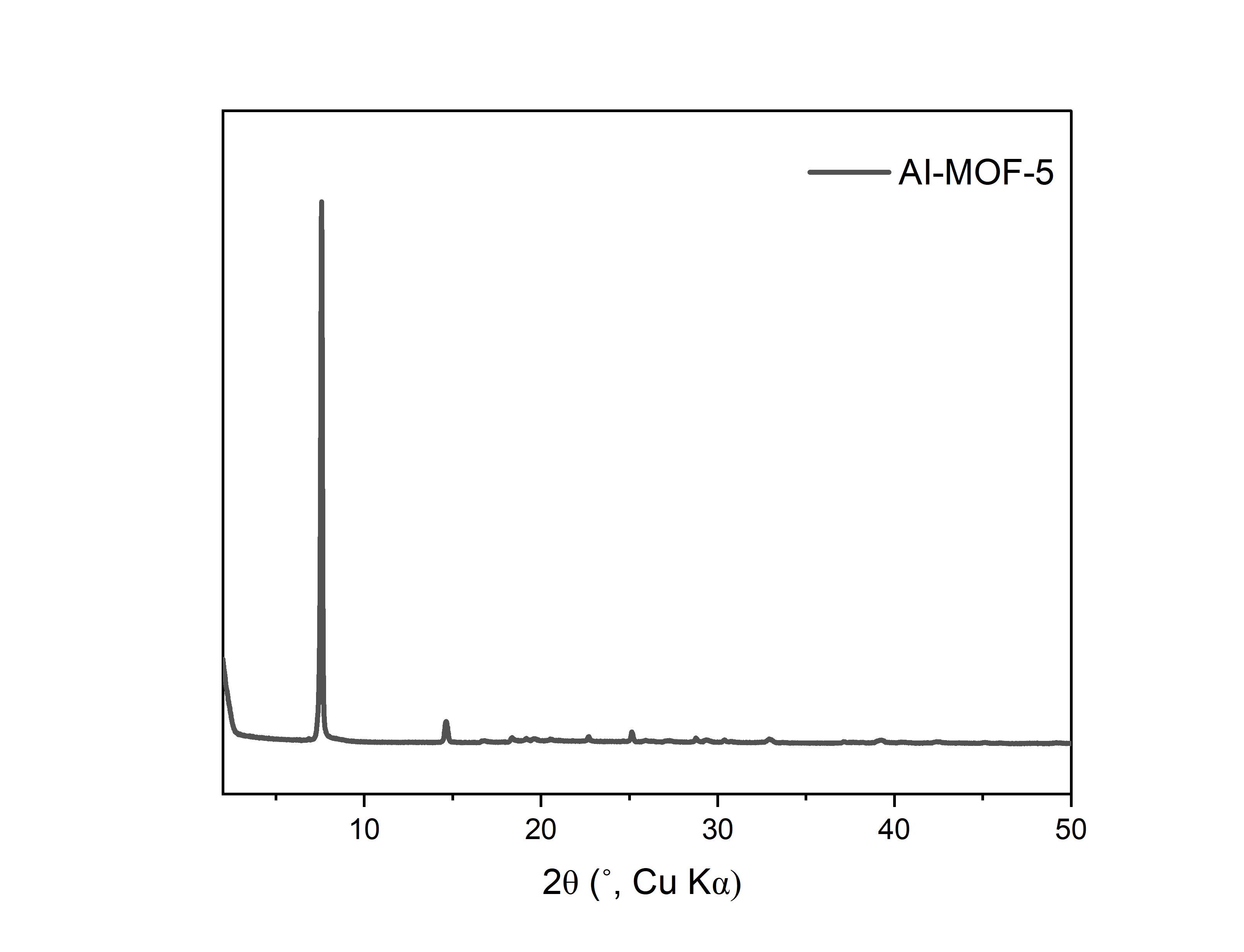}
  \caption{\textbf{Experimental PXRD of the aluminum dimethyl fumarate MOF with DMF solvent, AI-MOF-5.}}
  \label{fig:synth_AI-MOF-5_pxrd}
\end{figure}


\begin{table}
\centering
\caption{Benchmark comparison between GFN1-xTB and GFN2-xTB for MOF geometrical relaxation. Metrics include convergence success rate, computational time, and average deviations in lattice parameters and atomic positions after relaxation.}
\label{tabSI:GFN-xTB_analysis}
\begin{tabular}{lcc}
\hline
\textbf{Property} & \textbf{GFN1-xTB} & \textbf{GFN2-xTB} \\
\hline
Time [s] & 200.6 & 165.5 \\
Success Rate [\%] & 71.20 & 60.80 \\
Lattice Lengths [\AA{}] & 0.092 & 0.057 \\
Lattice Angles [$^{\circ}$] & 0.45 & 0.38 \\
Atom positions [\AA{}] & 0.2024 & 0.1677 \\
\hline
\end{tabular}
\end{table}

\begin{table}[h!]
\centering
\label{tabSI:bulk_modulus}
\caption{Mean and standard deviation ($\mu \pm \sigma$) of thermal decomposition temperature (T\textsubscript{d}), solvent removal stability, and MACE-MP-0b computed bulk modulus across QMOF, QMOF's subset and diffusion-generated, called ``Diffusion + LLM MOFs.''}
\begin{tabular}{lccc}
\hline
\textbf{Property} & \textbf{QMOFs} & \textbf{QMOF subset} & \textbf{Diffusion + LLM} \\
\hline
T\textsubscript{d} [$^\circ$C] & $331.3 \pm 55.5$ & $336.4 \pm 40.0$ & $325.6 \pm 16.8$ \\
Solvent removal Stability & $0.62 \pm 0.26$ & $0.65 \pm 0.24$ & $0.88 \pm 0.10$ \\
Bulk Modulus [GPa] & $16.5 \pm 7.9$ & $8.3 \pm 4.2$ & $3.7 \pm 1.5$ \\
\hline
\end{tabular}
\end{table}

\begin{table}
\centering
\caption{Crystallographic data for compounds SM\_621B. $^aR1 = \left[\frac{\sum w(F_o - F_c)^2}{\sum wF_o^2}\right]^{1/2}$; \quad
$^bwR2 = \left[\frac{\sum w(F_o^2 - F_c^2)^2}{\sum wF_o^2}\right]^{1/2}$,\\[5pt]
with \quad $w = \frac{1}{\sigma^2(F_o^2) + (aP)^2 + bP}$, \quad where \quad $P = \frac{\max(F_o^2,0) + 2F_c^2}{3}$.}
\label{tabSI:SM621B}
\begin{tabular}{ll}
\hline
\multicolumn{2}{c}{SM\_621B} \\
\hline
CCDC Number           & 2433335 \\
Chemical formula      & C$_{18}$H$_{18}$O$_{13}$Zn$_4\cdot$2(DMF) \\
Formula weight        & 849.99 \\
Space group           & Pna21 \\
$a$ (Å)               & 22.0122(10) \\
$b$ (Å)               & 18.1336(8) \\
$c$ (Å)               & 10.5378(5) \\
$\alpha$ (deg)        & 90 \\
$\beta$ (deg)         & 90 \\
$\gamma$ (deg)        & 90 \\
$V$ (Å$^3$)           & 4206.3(3) \\
$Z$                   & 4 \\
$\mu$ (mm$^{-1}$)     & 3.066 \\
$T$ (K)               & 200 \\
$R1^a$ ($wR2^b$)      & 0.0221 (0.0571) \\
Reflections collected & 26755 \\
Goodness-of-fit on $F^2$ & 1.100 \\
Radiation type        & Cu K$\alpha$ \\
\hline
\end{tabular}
\end{table}


\clearpage 






\end{document}